\documentclass[12pt,aps,amsmath,amssymb,nofootinbib,prd,superscriptaddress]{revtex4}
\usepackage{bm}
\usepackage{graphicx}
\usepackage{amsmath}
\usepackage{amsfonts,amsbsy}
\usepackage{amssymb}
\usepackage{placeins}
\usepackage[utf8]{inputenc}
\usepackage{color}
\usepackage[breaklinks, colorlinks=true,citecolor={blue}]{hyperref}

\def\gsim{ \,\, \vcenter{\hbox{$\buildrel{\displaystyle >}\over\sim$}}
 \,\,}
\def\lsim{ \,\, \vcenter{\hbox{$\buildrel{\displaystyle <}\over\sim$}}
 \,\,}
\def\be{\begin{equation}}
\def\ee{\end{equation}}
\def\bea{\begin{eqnarray}}
\def\eea{\end{eqnarray}}

\definecolor{dgreen}{cmyk}{1.,0.,1.,0.2}        
\definecolor{orange}{cmyk}{0.,0.353,1.,0.}    

\begin{document}

\title{Measuring the Weizs\"acker-Williams distribution of linearly
  polarized gluons at an EIC through dijet azimuthal asymmetries}

\author{Adrian Dumitru}
\affiliation{Department of Natural Sciences, Baruch College, CUNY,
17 Lexington Avenue, New York, NY 10010, USA}
\affiliation{The Graduate School and University Center, The City
  University of New York, 365 Fifth Avenue, New York, NY 10016, USA}

\author{Vladimir Skokov}
\affiliation{Department of Physics, North Carolina State University, Raleigh, NC 27695, USA}
\affiliation{RIKEN/BNL Research Center, Brookhaven National
  Laboratory, Upton, NY 11973, USA}

\author{Thomas Ullrich}
\affiliation{Brookhaven National
  Laboratory, Upton, NY 11973, USA}
\affiliation{Yale University, New Haven, CT 06520, USA}

\begin{abstract}
The production of a hard dijet with small transverse momentum
imbalance in semi-inclusive DIS probes the conventional and linearly
polarized Weizs\"acker-Williams (WW) Transverse Momentum Dependent (TMD)
gluon distributions. The latter, in particular, gives rise to an
azimuthal dependence of the dijet cross-section. In this paper we
analyze the feasibility of a measurement of these TMDs through dijet
production in DIS on a nucleus at an Electron-Ion Collider. We
introduce the MCDijet Monte-Carlo generator to sample quark --
antiquark dijet configurations based on leading order parton level
cross-sections with WW gluon distributions that solve the non-linear
small-$x$ QCD evolution equations. These configurations are fragmented
to hadrons using PYTHIA, and final state jets are reconstructed.
 We report on background studies and on the effect of
kinematic cuts introduced to remove beam jet remnants. We estimate
that with an integrated luminosity of 20 fb$^{-1} / A$ one can
determine the distribution of linearly polarized gluons with a
statistical accuracy of approximately 5\%.
\end{abstract}

\date{\today}

\maketitle

\newpage

\tableofcontents

\newpage

\section{Introduction}
\label{sec:intro}

Building an Electron-Ion Collider (EIC) is one of the key projects
of the nuclear physics community in the U.S. The main purpose of an
EIC is to study the gluon fields of QCD and provide
insight into the regime of non-linear color field
dynamics~\cite{Accardi:2012qut,Boer:2011fh}. The energy dependence of
various key measurements has been assessed recently in
Ref.~\cite{Aschenauer:2017jsk}.

In this paper we focus on the small-$x$ regime of strong color fields
in hadrons and nuclei~\cite{Mueller:1999wm}. An EIC, in principle, is
capable of providing clean measurements of a variety of correlators
of the gluon field in this regime. Here, we are interested, in
particular, in the conventional and linearly polarized
Weizs\"acker-Williams (WW) gluon distributions at small
$x$~\cite{Dominguez:2011wm,Dominguez:2011br}. These distributions
arise also in Transverse Momentum Dependent (TMD)
factorization~\cite{Mulders:2000sh, Bomhof:2006dp,
  Meissner:2007rx}. (For a recent review of TMD gluon distributions at
small $x$ see Ref.~\cite{Petreska:2018cbf}.)  Our main goal is to
conduct a first assessment of the feasibility of a measurement of
these gluon distributions at an EIC through the dijet production
process.

The WW
 TMD gluon distributions, and in particular the distribution of
 linearly polarized gluons, appears in a variety of processes. This
 includes production of a dijet or heavy quark pair in hadronic
 collisions~\cite{Boer:2009nc, Akcakaya:2012si, Marquet:2017xwy} or
 DIS at moderate~\cite{Boer:2010zf, Pisano:2013cya, Boer:2016fqd,
   Efremov:2017iwh, Efremov:2018myn} or high
 energies~\cite{Dominguez:2011wm, Dominguez:2011br, Dumitru:2015gaa}
 where the dependence on the dijet imbalance is explicitly present.
 Dijet studies
 are the main focus of this paper.  The WW gluon distributions
 could also be measured in photon pair~\cite{Qiu:2011ai}, muon
 pair~\cite{Pisano:2013cya}, quarkonium~\cite{Boer:2016bfj},
 quarkonium pair~\cite{Lansberg:2017dzg}, or quarkonium plus
 dilepton~\cite{Lansberg:2017tlc} production in hadronic
 collisions. The distributions also determine fluctuations of the divergence of the
 Chern-Simons current at the initial time of a relativistic heavy-ion
 collision~\cite{Lappi:2017skr}. Finally, we illustrate that the
 conventional WW gluon distribution at small $x$ could, in principle,
 be determined also from dijet production in ultraperipheral p+p, p+A,
 and A+A collisions. However, as explained in the next section, the
 distribution of linearly polarized gluons cannot be accessed with
 quasi real photons. This underscores the importance of conducting the
 dijet measurements at an EIC.

\section{Dijets in DIS at high energies}
\label{sec:dijet}

At leading order in $\alpha_s$ the cross-section for inclusive
production of a $q+\bar q$ dijet in high energy deep inelastic
scattering of a virtual photon $\gamma^*$ off a proton or nucleus is
given by~\cite{Metz:2011wb,Dominguez:2011wm}
\begin{eqnarray}
E_1E_2
\frac{d\sigma ^{\gamma _{T}^{\ast }A\rightarrow
    q\bar{q}X}}{d^3k_1d^3k_2 d^2b}
&=&\alpha _{em}e_{q}^{2}\alpha _{s}\delta \left( x_{\gamma ^{\ast
}}-z-\bar z\right) z\bar z\left( z^{2}+\bar z^{2}\right) \frac{\epsilon _{f}^{4}+
{P}_{\perp }^{4}}{({P}_{\perp }^{2}+\epsilon _{f}^{2})^{4}}
\notag \\
&&
\quad \quad \quad \quad \quad \quad
\times \left[ xG^{(1)}(x,q_{\perp })-\frac{2\epsilon _{f}^{2}{P}%
_{\perp }^{2}}{\epsilon _{f}^{4}+{P}_{\perp }^{4}}\cos
  \left(2\phi\right)xh_{\perp }^{(1)}(x,q_{\perp })\right] ~,
\label{eq:dijet_T} \\
E_1E_2
\frac{d\sigma ^{\gamma _{L}^{\ast }A\rightarrow q\bar{q}X}}{d^3k_1d^3k_2 d^2b}
&=&\alpha _{em}e_{q}^{2}\alpha _{s}\delta \left( x_{\gamma ^{\ast
}}-z-\bar z\right) z^2\bar z^2\frac{8\epsilon _{f}^{2}{P}_{\perp }^{2}}{(
{P}_{\perp }^{2}+\epsilon _{f}^{2})^{4}}  \notag \\
&&
\quad \quad \quad \quad \quad \quad
\times \left[ xG^{(1)}(x,q_{\perp })+\cos \left(2
  \phi\right)xh_{\perp }^{(1)}(x,q_{\perp })\right]~.
\label{eq:dijet_L}
\end{eqnarray}
Here, $x_{\gamma ^{\ast}}=1$, and
\be
\vec{P}_{\perp } = \bar z \vec{k}_{1\perp} - z \vec{k}_{2\perp}~~,~~
\vec q_\perp = \vec{k}_{1\perp}+\vec{k}_{2\perp}
\label{eq:Ptqtdef}
\ee
are the dijet transverse momentum (hard) scale $\vec{P}_{\perp }$ and
the momentum imbalance $\vec{q}_\perp$, respectively~\footnote{Here
  and below the transverse two dimensional component of a three
  dimensional vector $\vec{k} = (\vec{k}_\perp, k_z)$ are denoted by
  $\vec{k}_\perp$.}. Note that the momentum imbalance is explicitly preserved,
   enabling us to probe a regime of high gluon densities
at small $q_\perp$ even if $Q$ and $P_\perp$ exceed the
so-called gluon saturation scale $Q_s(x)$ at the given
$x$~\cite{Kovchegov:2012mbw}.

The transverse momenta of the produced quark and anti-quark are given
by $\vec{k}_{1\perp}$ and $\vec{k}_{2\perp}$ and their respective
light-cone momentum fractions are $z$ and $\bar z$. The invariant mass
of the dijet is $M_{\rm inv}=P_\perp/\sqrt{z \bar z}$;  for
massless quarks we have $\epsilon_f^2=z\bar z Q^2$.  We restrict our
consideration to the case when $\vec{P}_{\perp }$ is
greater than $\vec{q}_\perp$, also known as the ``correlation limit''~\cite{Dominguez:2011wm,Dominguez:2011br}.
The above
equations are valid to leading power in $1 / P_\perp^2$.  Power
corrections were derived in Ref.~\cite{Dumitru:2016jku}. They generate
corrections $\sim (Q_s^2/P_\perp^2) \log P_\perp$ to the isotropic and
$\sim \cos 2\phi$ terms detailed  above.  Moreover, a $\sim \cos 4\phi$
angular dependence arises from power corrections of order
$q_\perp^2/P_\perp^2$.

In Eqs.~(\ref{eq:dijet_T},\ref{eq:dijet_L}), $\phi$ denotes the azimuthal
angle between $\vec{P}_{\perp }$ and $\vec{q}_{\perp }$.
Note that we work in a frame  where neither the
virtual photon nor the hadronic target carries non-zero transverse
momentum before their interaction. For our jet reconstruction analysis
we  transform every event to such a frame.

The average $\cos 2\phi$ measures the azimuthal anisotropy,
\be \label{eq:Def_v2_phi}
v_2 \equiv \left< \cos 2 \phi\right>~.
\ee
The brackets denote an average over $\phi$ of $\cos 2 \phi$ at {\em
  fixed} $q_\perp$ and $P_\perp$, with
normalized weights proportional to the cross-sections in
Eqs.~(\ref{eq:dijet_T}) or~(\ref{eq:dijet_L}), respectively.

Since\footnote{$W$ in Eq.~(\ref{eq:x_g}) denotes the CM energy
of the $\gamma^*$ - nucleon collision.}
\be \label{eq:x_g}
x = \frac{1}{W^2+Q^2-M^2}\left( Q^2 +
q_\perp^2 + \frac{1}{z\bar z}P_\perp^2\right)
\ee
is independent of $\phi$, for definite polarization of the virtual
photon we have~\cite{Dumitru:2015gaa}
\be \label{eq:v2_L_T}
v_2^L = \frac{1}{2} \frac{xh_{\perp }^{(1)}(x,q_{\perp
  })}{xG^{(1)}(x,q_{\perp })}~~~,~~~
v_2^T = - \frac{\epsilon _{f}^{2}{P}_{\perp }^{2}}{\epsilon
  _{f}^{4}+{P}_{\perp }^{4}} \frac{xh_{\perp }^{(1)}(x,q_{\perp
  })}{xG^{(1)}(x,q_{\perp })}~.
\ee
The polarization determines the sign of $v_2$.  In experiments it is
not possible to tell the polarization of the photon in dijet
production directly. Instead, one measures the polarization blind sum,
see Eq.~\eqref{Eq:v2sum}.  In Sec.~\ref{sec:feasibility},  we
show how one could disentangle $v_2^L$ and $v_2^T$.

A measurement of the $\phi$-averaged dijet cross-section provides the
conventional (unpolarized) Weizs\"{a}cker-Williams gluon distribution
$xG^{(1)}(x,q^2_\perp)$ via
Eqs.~(\ref{eq:dijet_T},\ref{eq:dijet_L}). A measurement of the average
of $\cos 2\phi$ then provides the distribution of linearly polarized
gluons via Eqs.~(\ref{eq:v2_L_T}). We note that the conventional
distribution can, in principle, be measured in $\gamma A \to q\bar q
X$ also in the $Q^2\to0$ limit. However, for a real photon
$\epsilon_f^2 \propto Q^2 \to0$ so that the cross-section for the
process becomes isotropic and one no longer has access to $xh_{\perp
}^{(1)}(x,q^2_\perp)$.

Eqs.~(\ref{eq:dijet_T},\ref{eq:dijet_L}) are restricted to high
energies not only because the large component of the light cone
momenta of the quark and anti-quark are conserved (high-energy
kinematics), but also because we neglect photon - quark scattering
with gluon emission ($\gamma^* q \to g+q$). For an unpolarized target,
and massless quarks, the distribution $f_1^q(x,q^2)$ of unpolarized
quarks enters~\cite{Pisano:2013cya, Kovchegov:2015zha} and gives an
additional contribution to the isotropic part of the dijet
cross-section. For more realistic computations at EIC energies these
contributions should be included in the future.

The linearly polarized and conventional gluon distributions\footnote{We
  only consider the {\em forward} gluon distributions in this
  paper. In the non-forward case the general decomposition of the WW
  GTMD involves additional independent functions on the r.h.s.\ of
  Eq.~(\ref{eq:xGij_xG_xh}), see e.g. Ref.~\cite{Boussarie:2018zwg}.}
are given by the traceless part and by the trace of the
Weizs\"{a}cker-Williams unintegrated gluon distribution, respectively:
\begin{equation}
xG^{ij}_{\rm WW}(x,q_{\perp }) = \frac{1}{2} \delta^{ij}
xG^{(1)}(x,q_{\perp }) -\frac{1}{2} \left(\delta^{ij} - 2
\frac{q^i_\perp q^j_\perp}{q^2_\perp} \right)
xh^{(1)}_\perp(x,q_{\perp }) ~.  \label{eq:xGij_xG_xh}
\end{equation}
Their general operator definitions in QCD were provided in
Refs.~\cite{Mulders:2000sh,Bomhof:2006dp,Meissner:2007rx}.  At small $x$, $x G^{ij}_{\rm WW}(x,q^2_{\perp })$ is
expressed as a two-point correlator of the field in $A^+=0$ light cone
gauge~\cite{Kharzeev:2003wz, Dominguez:2011wm, Dominguez:2011br}:
\be
\alpha_s\, x G^{ij}_{\rm WW} (x, q_\perp) =
\frac{2}{S_\perp} \int
\frac{d^2 x_\perp} {(2\pi)^2}
\frac{d^2 y_\perp} {(2\pi)^2}\,
e^{-i \vec{q}_\perp \cdot (\vec{x}_\perp-\vec{y}_\perp)}\,
\left\langle
gA^{i,a}(\vec x_\perp) \,\, gA^{j,a}(\vec y_\perp)
\right\rangle\,.
\label{eq:Def_Gij_smallx}
\ee
Here $S_\perp$ denotes the transverse area of the target and
$gA^i(\vec{x}_\perp) = -iU^\dagger (\vec{x}_\perp) \partial^i U
(\vec{x}_\perp)$, with the conventional definition of the Wilson line
in the fundamental representation, $U (\vec{x}_\perp)$.
$\langle\cdots\rangle$ in Eq.~(\ref{eq:Def_Gij_smallx}) refers to an
average over all quasi-classical configurations of small-$x$ gluon
fields.  At small $x$ the function $\left(2q^i_\perp
q^j_\perp/q^2_\perp -\delta^{ij}\right) A^i(\vec q_\perp)A^j(-\vec
q_\perp)$ exhibits large fluctuations across configurations, in
particular for $q_\perp$ not too far above the saturation scale
$Q_s$~\cite{Dumitru:2017ftq}. However, in the single dijet production
process one can only determine the average $xh_{\perp
}^{(1)}(x,q^2_{\perp })$.

\begin{figure}
\centering
\includegraphics[width=0.45\linewidth]{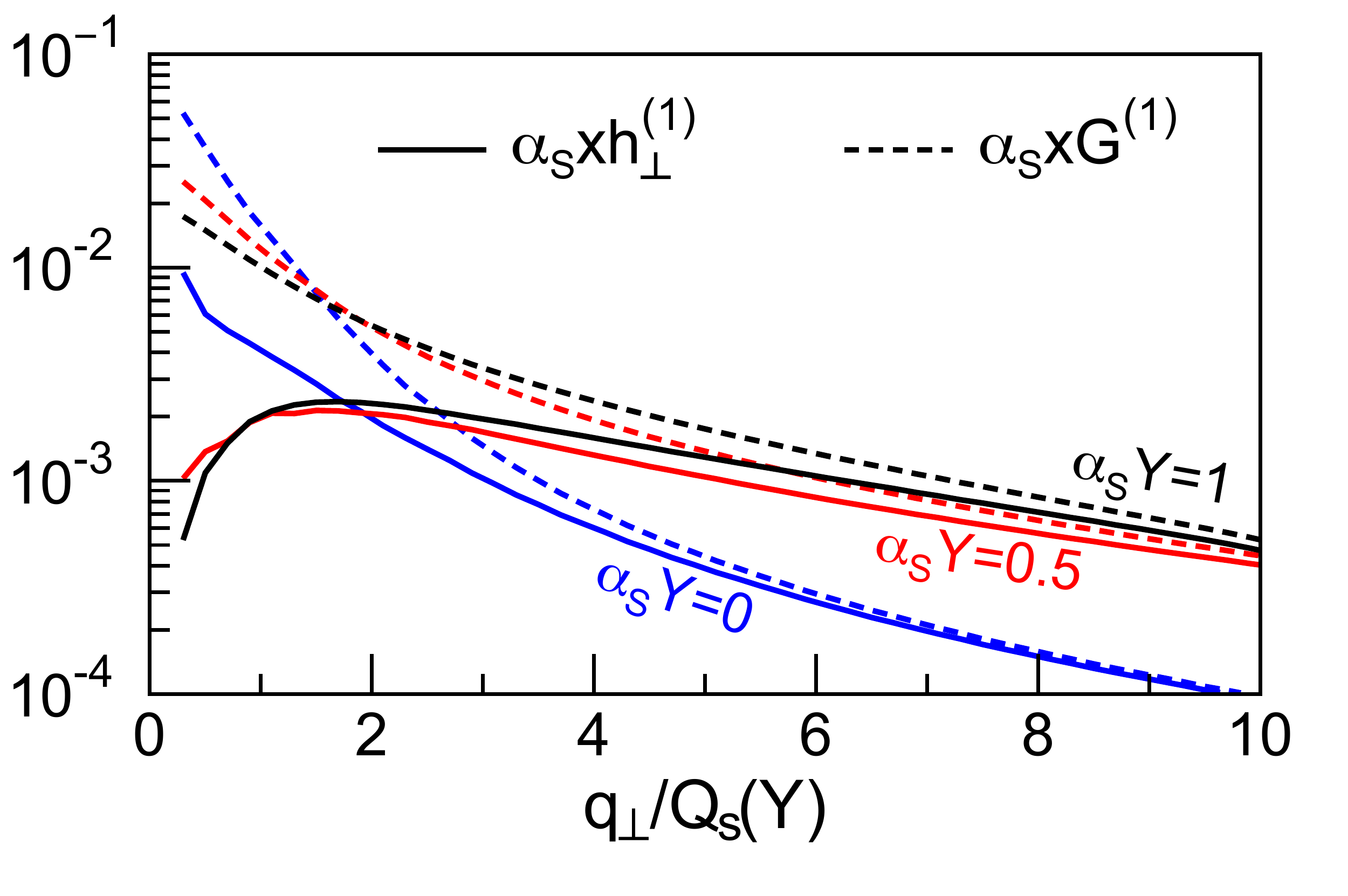}
\includegraphics[width=0.45\linewidth]{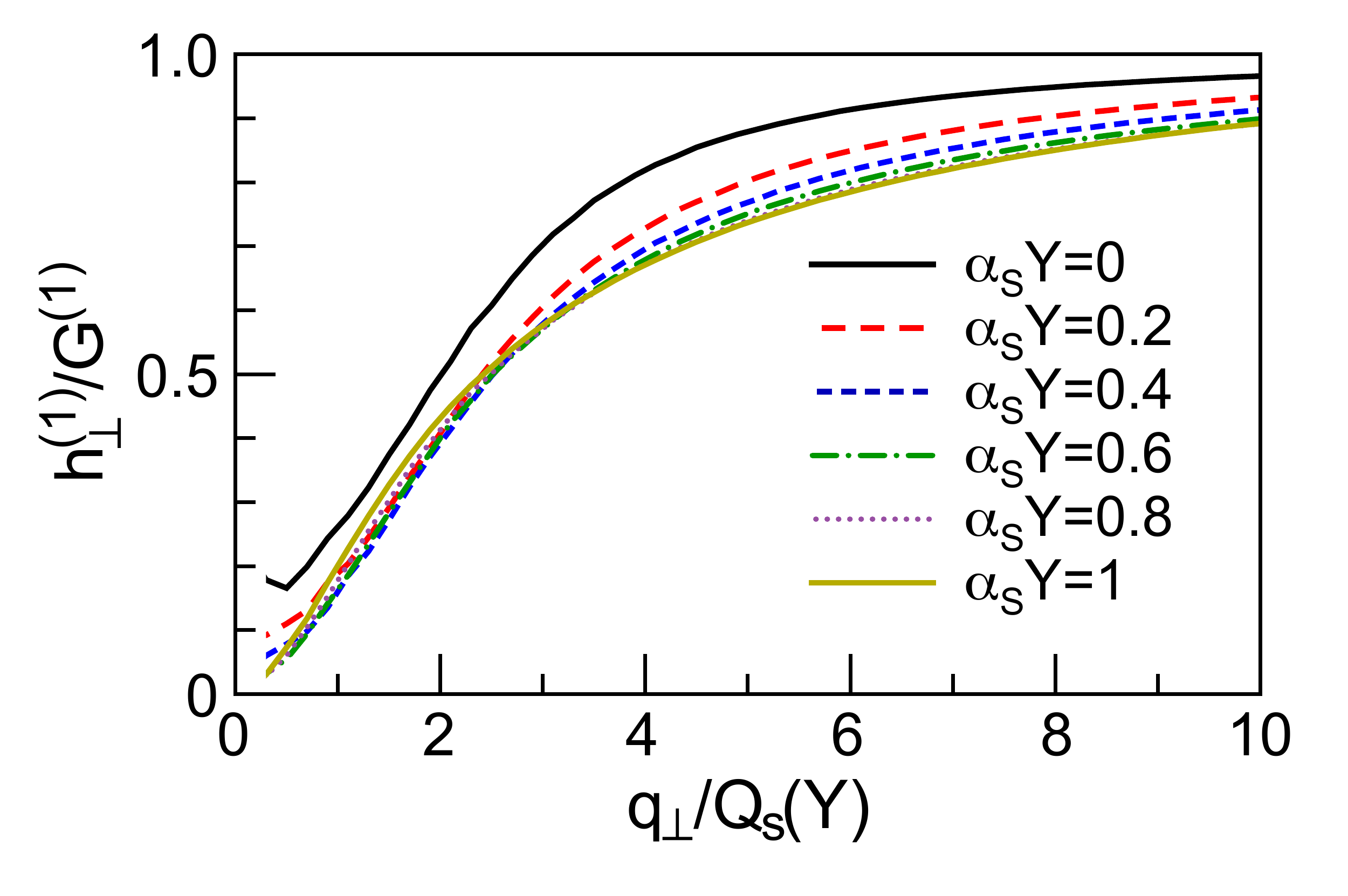}
\caption{$xG^{(1)}(x,q^2_\perp)$ and $xh^{(1)}(x,q^2_\perp)$ WW gluon
  distributions versus transverse momentum $q_\perp$ at different rapidities
  $Y=\log x_0/x$. $Q_s(Y)$ is the saturation momentum. The curves correspond to
  evolution at fixed $\alpha_s$~\cite{Dumitru:2015gaa}.}
\label{fig:xGxh}       
\end{figure}
The functions $xG^{(1)}(x,q^2_{\perp })$ and $xh_{\perp
}^{(1)}(x,q^2_{\perp })$ for the McLerran-Venugopalan (MV)
model~\cite{McLerran:1994ni, McLerran:1994ka} of a large nucleus were computed analytically in Refs.~\cite{Metz:2011wb,Dominguez:2011br}.
Explicit expressions for a more general theory of
Gaussian fluctuations of the covariant gauge field $A^+$ were
given in Ref.~\cite{Dumitru:2016jku}; also see
Refs.~\cite{Marquet:2016cgx,Albacete:2018ruq}. Numerical solutions of
the JIMWLK evolution equations~\cite{Balitsky:1995ub, Balitsky:1998kc,
  Balitsky:1998ya, Jalilian-Marian:1997jx, Jalilian-Marian:1997gr,
  JalilianMarian:1997dw, Kovner:1999bj, Kovner:2000pt, Iancu:2000hn,
  Iancu:2001ad, Ferreiro:2001qy, Weigert:2000gi} to small $x$ were presented in Refs.~\cite{Dumitru:2015gaa, Marquet:2016cgx}, shown
in Fig.~\ref{fig:xGxh}. At high transverse momentum one finds that
$xh^{(1)}(x,q^2_\perp)\to xG^{(1)}(x,q^2_\perp)$ corresponding to
maximal polarization.  On the other hand, at low $q_\perp$ one has
$xh^{(1)}(x,q^2_\perp)/ xG^{(1)}(x,q^2_\perp)\ll 1$, implying that
there the angular dependence of the cross-section~(\ref{eq:dijet_T},
\ref{eq:dijet_L}) is weak. For $q_\perp\sim Q_s(Y)$ these numerical
solutions predict a substantial angular modulation of the dijet cross-section since $xh^{(1)}(x,q^2_\perp)/ xG^{(1)}(x,q^2_\perp)\simeq 10\%
- 20\%$.

Our event generator described in the following
Sec.~\ref{sec:mcdijet} employs tabulated solutions of the leading
order, fixed coupling JIMWLK evolution
equations~\cite{Balitsky:1995ub, Balitsky:1998kc, Balitsky:1998ya,
  Jalilian-Marian:1997jx, Jalilian-Marian:1997gr,
  JalilianMarian:1997dw, Kovner:1999bj, Kovner:2000pt, Iancu:2000hn,
  Iancu:2001ad, Ferreiro:2001qy, Weigert:2000gi} for
$xh^{(1)}(Y,q^2_\perp)$ and $xG^{(1)}(Y,q^2_\perp)$, where $Y=\log
x_0/x$. The initial condition at $x_0=0.01$ is given by the MV
model. In particular, the initial MV saturation scale is set to
$Q_s(x_0)= 1.2$~GeV corresponding to a large nucleus with $A\sim 200$
nucleons, on average over impact parameters.

\subsection{Moments of inter-jet azimuthal angle}
\label{sec:cos2Phi}

In this subsection we discuss the relation of $\langle\cos
2\phi \rangle$ introduced in the previous section to $\langle\cos
2\Phi \rangle$, where $\Phi$ is the azimuthal angle between the two
jets (i.e.\ between $\vec k_{1\perp}$ and $\vec k_{2\perp}$). They are related
through
\be
\cos^2 \Phi = \frac{(\vec k_{1\perp}\cdot \vec
  k_{2\perp})^2}{k_{1\perp}^2\, k_{2\perp}^2}~.
\ee
To obtain moments in the correlation limit at {\em fixed} $q_\perp$
and $P_\perp$ one inverts equations~(\ref{eq:Ptqtdef}) to express
$\vec k_{1\perp}=\vec P_\perp + z \vec q_\perp$ and $\vec k_{2\perp}=-
\vec P_\perp + \bar z  \vec q_\perp$, and performs an expansion of
$\cos^2 \Phi$ in powers of $q_\perp / P_\perp$. This leads to
\bea
\cos 2\Phi &=& 2\cos^2\Phi-1 = 1 + \frac{q_\perp^2}{P_\perp^2}\left( \cos 2\phi
-1\right) \nonumber\\
    & & + \frac{q_\perp^4}{P_\perp^4}\left[ z \bar z -
\left(1-2z\bar z \right)\cos 2\phi + \left(1-3z\bar z \right)\cos 4\phi\right] \nonumber\\
& & + \cdots ~.
\eea
We have dropped terms which vanish upon integration over $\phi$. The
dots indicate contributions of higher order in $q_\perp/P_\perp$.
Taking an average\footnote{Recall that this average is performed with
  normalized weights $w_{L,T}(\phi)$ proportional to the cross-sections~(\ref{eq:dijet_T},\ref{eq:dijet_L}), respectively.} over
$\phi$ at fixed $q_\perp$ and $P_\perp$ we obtain
\bea
\left<\cos 2\Phi\right>\bigr|_{q_\perp, P_\perp} &=& 1 +
\frac{q_\perp^2}{P_\perp^2}\left< \cos 2\phi -1\right>  \nonumber\\
    & & + \frac{q_\perp^4}{P_\perp^4}\left[ z \bar z -
\left(1-2z\bar z  \right)\left<\cos 2\phi\right> + \left(1-3z\bar z \right)\left<\cos 4\phi\right>\right] \nonumber\\
& & + \cdots ~.
\label{eq:<cos2Phi>}
\eea
Since $d^2k_{1\perp}\, d^2k_{2\perp} \, \delta(q_\perp^2-(\vec k_{1\perp} +
\vec k_{2\perp})^2) \, \delta(P_\perp^2-(\bar z \vec k_{1\perp} -
z \vec k_{2\perp})^2) \sim d\phi$, the integral over $\phi$ is
equivalent to an integral over $\vec k_{1\perp}$ and $\vec k_{2\perp}$
at fixed $q_\perp$ and $P_\perp$. On the r.h.s.\ of
Eq.~(\ref{eq:<cos2Phi>}) one can now replace $\langle\cos
2\phi\rangle$ by $xh_{\perp }^{(1)}(x,q_{\perp }) \, / \,
xG^{(1)}(x,q_{\perp })$ times a prefactor, see
Eq.~(\ref{eq:v2_L_T}). Note that this ratio of gluon distributions
appears in $\left<\cos 2\Phi\right>$ with a suppression factor of
$q_\perp^2 / P_\perp^2$ whereas it contributes at ${\cal O}(1)$ to
$\left<\cos 2\phi\right>$. Moreover, $xh_{\perp }^{(1)}(x,q_{\perp }) \, / \,
xG^{(1)}(x,q_{\perp })$ also contributes at order $q_\perp^4 /
P_\perp^4$ while power corrections to $\left<\cos 2\phi\right>$ only
involve different correlators~\cite{Dumitru:2016jku}.

\subsection{Electron-proton/nucleus scattering}
\label{sec:eA}

The cross-section for dijet production in electron-nucleus scattering
is given by the product of the virtual photon fluxes of the electron
with the $\gamma^*$-nucleus cross-sections discussed
above~\cite{Budnev:1974de, Drees:1988pp, Nystrand:2004vn}:
\begin{equation}
  \frac{d\sigma^{e^-A \rightarrow q\bar{q}X}_{L,T}}{d Q^2 d W^2
    d^2P_\perp d^2q_\perp  dz}
  = f_{L,T}(Q^2, W^2)\,
  \frac{d \sigma^{\gamma _{L,T}^{\ast }A\rightarrow
    q\bar{q}X}}{d^2P_\perp d^2q_\perp dz}~.
\label{eq:sigmaLT_eA}
\end{equation}
Here,
\be
 \frac{d \sigma^{\gamma _{L,T}^{\ast }A\rightarrow
     q\bar{q}X}}{d^2P_\perp d^2q_\perp dz} =
 \int d^2b \int d\bar z \, E_1E_2 \frac{d\sigma ^{\gamma _{L,T}^{\ast }A\rightarrow
    q\bar{q}X}}{d^3k_1d^3k_2 d^2b}~.
\ee
The of transversely and longitudinally polarized photon fluxes
are given by
\bea
f_T (Q^2, W^2)  &=& \frac{\alpha_{\rm em}}{2 \pi Q^2 s y} \left( 1 +(1-y)^2 \right)~,
\label{eq:Fperp}\\
f_L (Q^2, W^2)  &=& \frac{\alpha_{\rm em}}{\pi Q^2 s y}  (1-y)~,
\label{eq:Flong}
\eea
with the inelasticity
\begin{equation}
y = \frac{W^2-M^2+Q^2}{s-M^2}~.
\label{eq:y}
\end{equation}
$M$ denotes the mass of the proton and $\sqrt{s}$ is the CM energy of
the $e^-$ - proton collision. The $\gamma^*$-proton/nucleus cross-section
on the r.h.s.\ of Eq.~(\ref{eq:sigmaLT_eA}) depends on $W^2$
and $Q^2$ through Eq.~(\ref{eq:x_g}).
Note that
Eqs.~(\ref{eq:Fperp},\ref{eq:Flong}) do not apply in the limit
 $Q^2\to0$
where the photon flux is effectively cut off at
$Q^2_\text{min}\sim \xi^2m_e^2/(1-\xi)$~\cite{Budnev:1974de}, with
$m_e$ the mass of the electron and $\xi$ the momentum fraction of the
photon relative to the electron. We are not concerned with $Q^2\lsim
1$~GeV$^2$ or $\xi\to1$ here and hence ignore the modification of
$f_{T,L}$ at low photon virtualities.

For given $W$ and $Q^2$, Bjorken-$x$ is
defined as
\be
x_\text{Bj} = \frac{Q^2}{ W^2 - M^2 + Q^2}~.
\label{eq:Bj_x}
\ee
%

\section{The event generator MCDijet}
\label{sec:mcdijet}

\subsection{General description}
The goal of the event generator MCDijet is to perform Monte-Carlo
sampling of the dijet (quark and anti-quark) production cross-section
described by Eq.~\eqref{eq:sigmaLT_eA}. The code is open source and
publicly available~\cite{git-MCDijet}.

In what follows, we will often refer to the acceptance-rejection
method (ACM) of generating random variables from a given probability
distribution; although this method is fairly basic, it nevertheless
proved sufficient for generating the required number of events on a
single processor in a reasonable amount of time.

In order to make the MC generator computationally feasible we have
adopted the following simplifying assumptions and approximations:
\begin{enumerate}
\item The dependence of the cross-section on the atomic number $A$ of
  the target enters via a single scale -- the saturation
  momentum\footnote{Throughout the manucsript we refer to the
    saturation scale for a dipole in the fundamental representation.},
  $Q_{s0}\sim A^{1/6}$, at $x=x_0=0.01$. For a Au nucleus, averaged
  over impact parameters, we assume that $Q_{s0}=1.2$~GeV. This is
  compatible with $Q_{s0}\approx 0.44$~GeV for a proton target
  extracted in Refs.~\cite{Albacete:2009fh,Albacete:2010sy} from fits
  to HERA data. The current implementation is restricted to impact
  parameter averaged dijet production; realistic nuclear
  thickness functions and fluctuations of the nucleon configurations
  in the nucleus have not been implemented.
\item The Wilson lines in the field of the target at $x=x_0$ are
  sampled using the MV model. They are then evolved to $x<x_0$
  using the fixed coupling Langevin form~\cite{Blaizot:2002np,Rummukainen:2003ns} of the JIMWLK renormalization
  group equation~\cite{Balitsky:1995ub,Balitsky:1998kc,Balitsky:1998ya,Jalilian-Marian:1997jx,Jalilian-Marian:1997gr,JalilianMarian:1997dw,Kovner:1999bj,Kovner:2000pt,Iancu:2000hn,Iancu:2001ad,Ferreiro:2001qy,Weigert:2000gi}, as described in
  Ref.~\cite{Dumitru:2014vka}. Note that for many phenomenological
  applications running coupling corrections are known to be important;
  they are neglected in the current version of the event
  generator. Also, the JIMWLK evolution ``time'' $t=\alpha_s Y$ is
  converted to a momentum fraction $x/x_0=\exp(-t/\alpha_s)$ using
  $\alpha_s=0.25$.
\item The Wilson lines are used to compute the dependence of
  $xG^{(1)}$ and $xh^{(1)}_\perp$ on the transverse momentum,
  $q_\perp$, and on $x$. The distributions are then averaged over
  the MV ensemble at the initial $x=x_0$, and over realizations of
  Langevin noise in small-$x$ evolution. The obtained averaged
  distributions are tabulated and stored in the file "misc.dat" which
  will be used by the MCDijet generator. We therefore do not propagate
  configuration by configuration fluctuations into actual
  event-by-event fluctuations in quark anti-quark production.
\end{enumerate}

MCDijet then performs the steps listed below:
\begin{itemize}
\item Using ACM based on the cross-section summed with respect to
  polarizations,
\begin{equation}
\frac{ d^2 \sigma_{L, T } (Q^2, W^2)}{d Q^2 d W^2}= f_{L, T}
(Q^2, W^2)\int d P_\perp d q_\perp dz d\phi \frac{d
  \sigma_{L,T}}{d P_\perp d q_\perp dz d\phi }~,
\label{Eq:MCsigmas}
\end{equation}
where the integration is performed in a restricted range of $P_\perp$
and $q_\perp$ specified below, we sample $Q^2$ and $W^2$ in the ranges
4 GeV$^2$ $< Q^2 < \frac{S-M^2}{1-x_0 }x_0$ and $M^2+Q^2\left(
\frac{1}{x_0}-1\right) < W^2 < s $.  The cross-sections $\frac{d
  \sigma_{L,T}}{d P_\perp d q_\perp dz d\phi }$ involve the WW
distribution functions and thus implicitly depend on $x$, given in
Eq.~(\ref{eq:x_g}).  Note that the calculation presented in this paper
are based on the leading order expressions~(\ref{eq:dijet_T},
\ref{eq:dijet_L}).  More realistic estimates of the absolute
cross-section may require a multiplicative K-factor $K>1$. Here we
provide a lower bound for the absolute cross-section and refrain from
using a K-factor.

\item The virtual photon may have either longitudinal or transverse
  polarization; it is assigned by sampling a random number $0<r<1$ uniformly.
  If
$$r<\frac{\sigma_{L } (Q^2, W^2)}{ \sigma_{L } (Q^2, W^2) +
    \sigma_{T } (Q^2, W^2)}$$  the polarization is
  longitudinal; otherwise it is transverse.

\item Using ACM and the differential cross-section for the photon
  polarization defined previously we generate a sample for
  $P_\perp$, $q_\perp$, $z$ and $\phi$.

\item Using the obtained $P_\perp$, $q_\perp$, $\phi$ and $z$, we can
  compute the transverse components of the quark ($k_1$) and anti-quark
  ($k_2$) momenta
\begin{align}
\vec{k}_{1\perp} & = P_\perp \vec{e}_P + z q_\perp \vec{e}_q \,
,  \label{eq:k1T_analysis-frame} \\
\vec{k}_{2\perp} & = -P_\perp \vec{e}_P + \bar{z} q_\perp \vec{e}_q\, ,
 \label{eq:k2T_analysis-frame}
\end{align}
where $\vec{e}_P = (\cos (\psi), \sin(\psi) )$ and $\vec{e}_q = (\cos
(\psi+\phi), \sin (\psi+\phi))$.  Here, $\psi$ is sampled
uniformly over $[0,2\pi[$.

\item Finally, the longitudinal momenta are given by
\begin{align}
k_{1 z} &= \frac{1}{\sqrt{2}} z q^+ - \frac{k_{1\perp}^2 } {2 \sqrt{2}
  z q^+}, \label{eq:k1z_analysis-frame}\\
  k_{2 z} &= \frac{1}{\sqrt{2}} \bar{z} q^+ -
  \frac{k_{2\perp}^2 } {2 \sqrt{2} \bar{z}  q^+}, \label{eq:k2z_analysis-frame}
\end{align}
where
\begin{align}
q^+ &= \sqrt{2} \, y'\, E_e; \quad \quad y' = \frac{1}{2}
y  \left(1 +  \sqrt{1+\left( \frac{2 x_{\rm Bj} M}{Q}\right)^2  }\right)~,
\label{eq:q+_analysis-frame}\\
q^- &= -\frac{Q^2}{2q^+}~,  \label{eq:q-_analysis-frame}
\end{align}
and $x_{\rm Bj}$ and $y$ are defined in Eqs.~\eqref{eq:Bj_x} and  \eqref{eq:y} respectively.
Here, $M$
denotes the mass of a proton, $E_e$ is the energy of the electron in
the lab frame, and $q^+$ is the large light-cone component of the
four-momentum $q^\mu$ of the virtual photon\footnote{Our convention
  here is that the longitudinal momentum of the virtual photon is
  positive. This is the most common convention in the theoretical
  literature.}.
\end{itemize}
The sampled kinematic variables and the corresponding numerical value
for the cross-section are then passed to Pythia. The interface between
Pythia and MCDijet is described in Sec.~\ref{sec:feasibility}.

\begin{figure}[t]
  \includegraphics[width=0.5\linewidth]{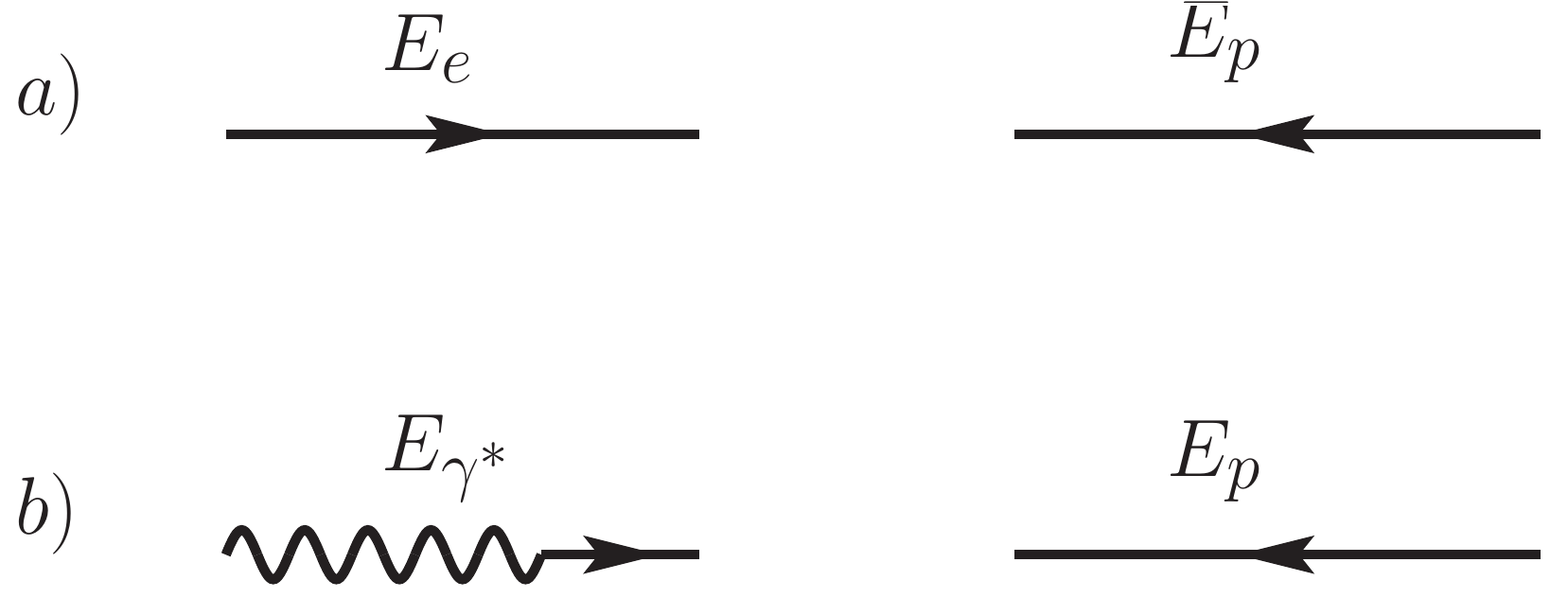}
  \caption{The reference frames: \\ a) The laboratory frame. In the
    laboratory frame, the electron and the proton have zero transverse
    momenta; the energy of the electron (proton) is $E_e$ ($E_p$).
    \\ b) The analysis frame. Here, the virtual photon and the proton
    have zero transverse momenta; the energy of the proton is the same
    as in the laboratory frame, equal to $E_p$. The energy of the
    virtual photon is $E_{\gamma^*} = (q^++q^-)/\sqrt{2}$, see
    Eqs.~\eqref{eq:q+_analysis-frame} and
    \eqref{eq:q-_analysis-frame}.  }
  \label{Fig:McDijet_ref_fr}
\end{figure}

The momentum assignments~(\ref{eq:k1T_analysis-frame} -
\ref{eq:q-_analysis-frame}) define the specific frame in which we
perform the analysis, see Fig.~\ref{Fig:McDijet_ref_fr}. That is, in
this frame the transverse momenta of the virtual photon and of the
target both vanish, the energy $E_p$ of the target nucleon(s) is equal
to that in the lab frame, and the invariant $\gamma^*$ - nucleon
collision energy squared is $W^2$. While, in principle, the analysis
could be performed in any other longitudinally boosted frame, such as
the Breit frame (see Appendix~\ref{sec:appB}) or the $\gamma^*$ -
nucleon center of momentum frame, we have found that the
reconstruction of the produced jets and of the target beam remnant is
rather accurate in this ``fixed $E_p$'' frame; see
Sec.~\ref{sec:feasibility} for further details.

\subsection{Numerical results}

\begin{figure}[htb]
  \centering {
    \includegraphics[width=0.45\linewidth]{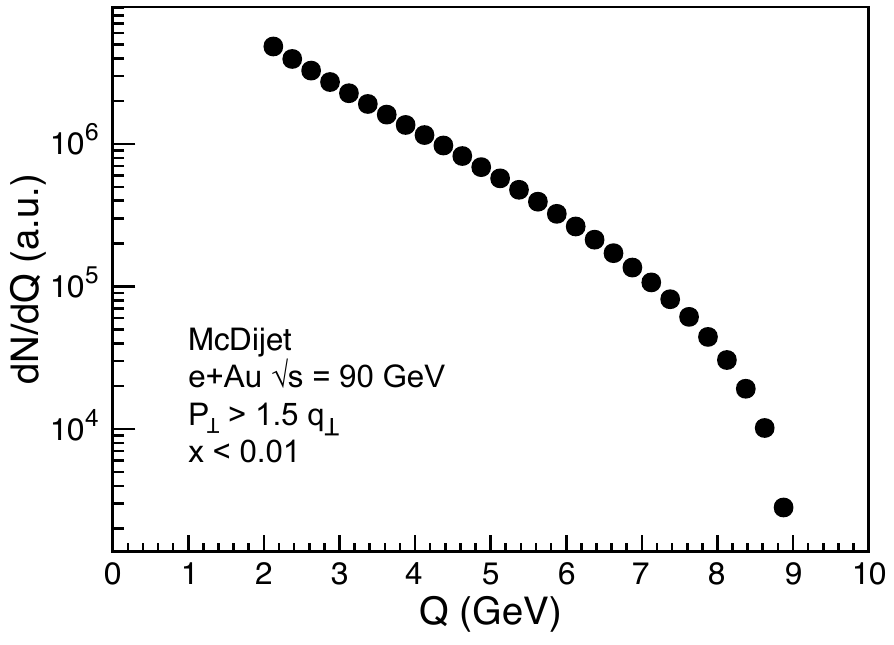}
    \includegraphics[width=0.45\linewidth]{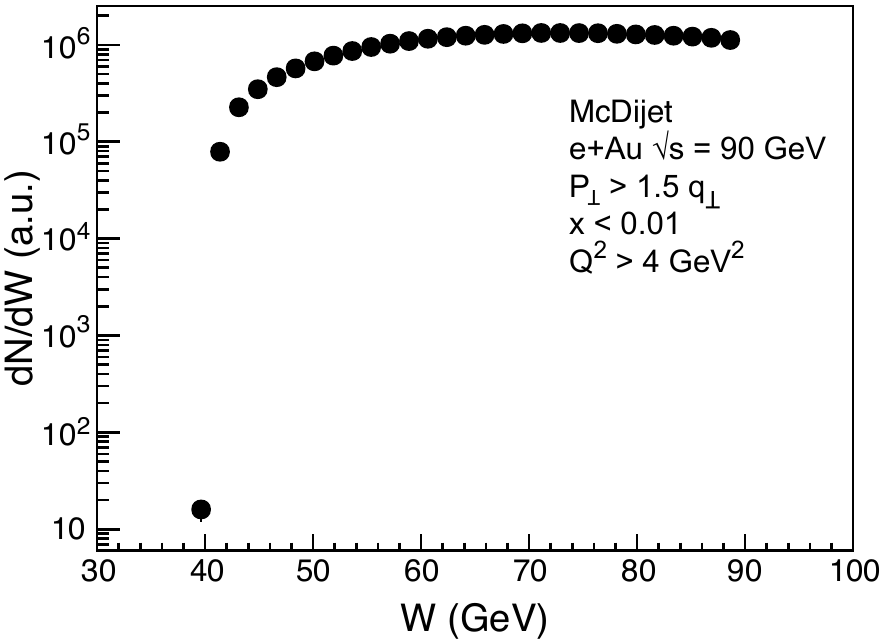}
    }
\caption{Distributions of photon virtuality $Q$ and
  $\gamma^*$-nucleon c.m.\ collision energy $W$ for dijet events
  subject to the kinematic cuts described in the text.}
	\label{fig:P(Q)_P(W)}
\end{figure}
\begin{figure}[htb]
  \centering {
    \includegraphics[width=0.45\linewidth]{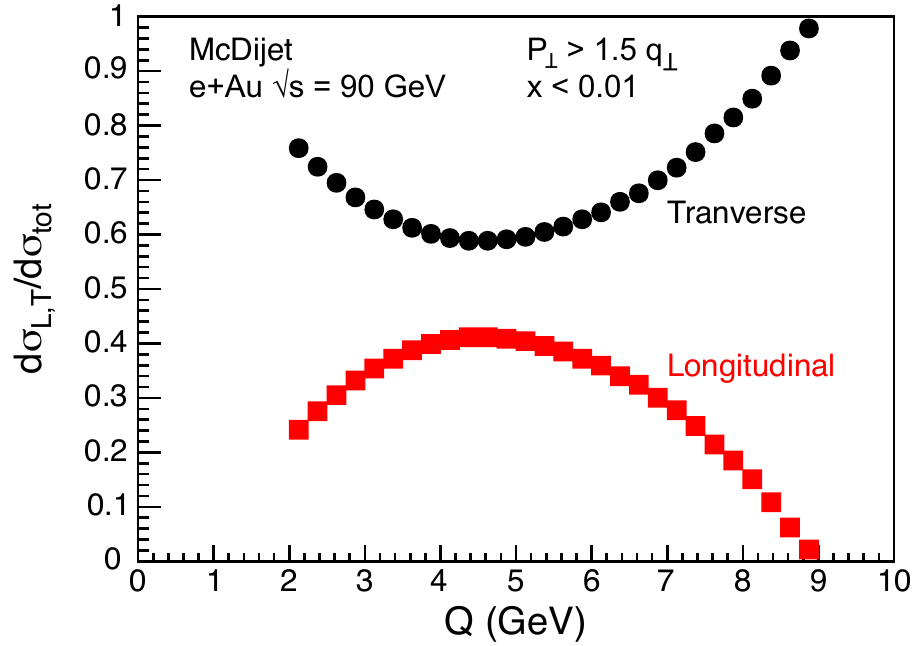}
    \includegraphics[width=0.45\linewidth]{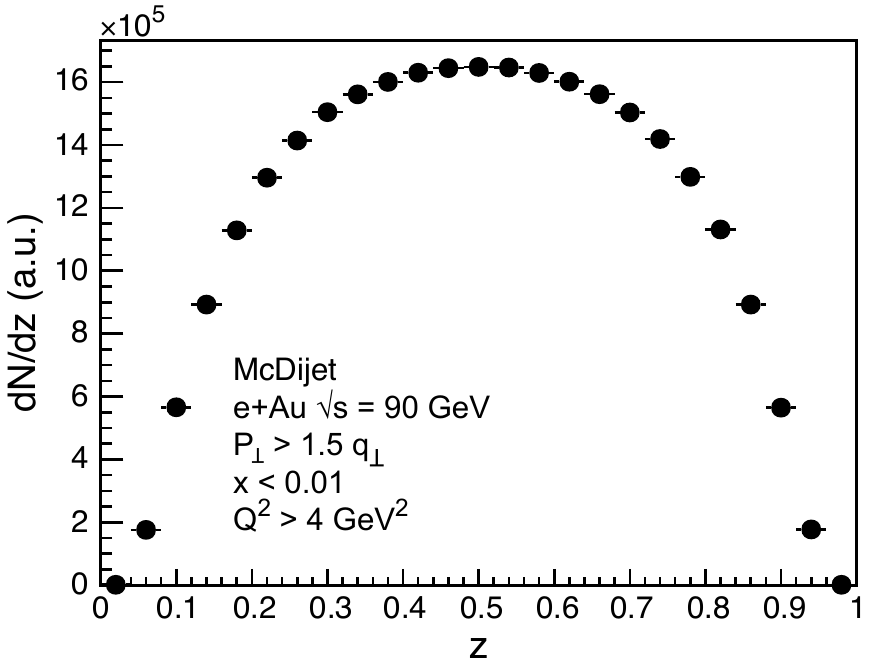}
    }
\caption{Left: The contributions of transverse vs.\ longitudinal photon
  polarizations as functions of $Q$. Right: The distribution of the quark
  momentum fraction $z$.}
	\label{fig:P(pol)_P(z)}
\end{figure}

In this subsection we show the distribution of dijet events over
various kinematic variables. The target is assumed to be Au with
$A=197$ nucleons, the $e^- - $Au collision energy is
$\sqrt{s}=90$~GeV. The event selection cuts are $\surd Q^2>2$~GeV,
$P_\perp > 1.5 \, q_\perp$, $q_\perp>1$~GeV, and $x$,
$x_\text{Bj}<0.01$. The distributions of $Q$ and $W$ are shown in
Fig.~\ref{fig:P(Q)_P(W)}, those of photon polarizations and quark
momentum fractions $z$ in Fig.~\ref{fig:P(pol)_P(z)}.

\section{Feasibility study for an Electron-Ion Collider}
\label{sec:feasibility}
In this section, based on the theoretical foundation outlined above,
we present a detailed study of the feasibility, requirements, and
expected precision of measurements of the azimuthal anisotropy of
dijets at a future Electron-Ion Collider (EIC).  We find that, at an
EIC \cite{Accardi:2012qut}, it is feasible to perform these
measurement although high energies, $\sqrt{s} \sim 100$ GeV, large
integrated luminosity of $\int L \mathrm{d}t \geq 20$ fb$^{-1}$, and
excellent jet capabilities of the detector(s) will be required.

In order to verify the feasibility we have to show that ({\em i}) the
anisotropy described by MCDijet (see Sec.~\ref{sec:mcdijet}) is
maintained in the reconstructed dijets measured in a realistic
detector environment, that ({\em ii}) the DIS background processes can
be suppressed sufficiently to not affect the level of anisotropy, and
({\em iii}) that $v_2^L$ and $v_2^T$ can be separated.

All studies presented here, were conducted with electron beams of
20~GeV and hadron beams with 100 GeV energy resulting in a
center-of-mass energy of $\sqrt{s} = 90$~GeV. As previously mentioned,
our convention is that the electron (hadron) beam has positive
(negative) longitudinal momentum. We use pseudo-data generated by the
Monte Carlo generator MCDijet, PYTHIA 8.2 \cite{Sjostrand:2014zea} for
showering of partons generated by MCDijet, and PYTHIA 6.4
\cite{Sjostrand:2006za} for background studies. Jets are reconstructed
with the widely used FastJet package~\cite{Cacciari:2011ma}.

\subsection{Azimuthal anisotropy of dijets} 
\label{sec:mcdijetStudy}
\begin{figure}[htb]
	\centering \includegraphics[width=0.85\linewidth]{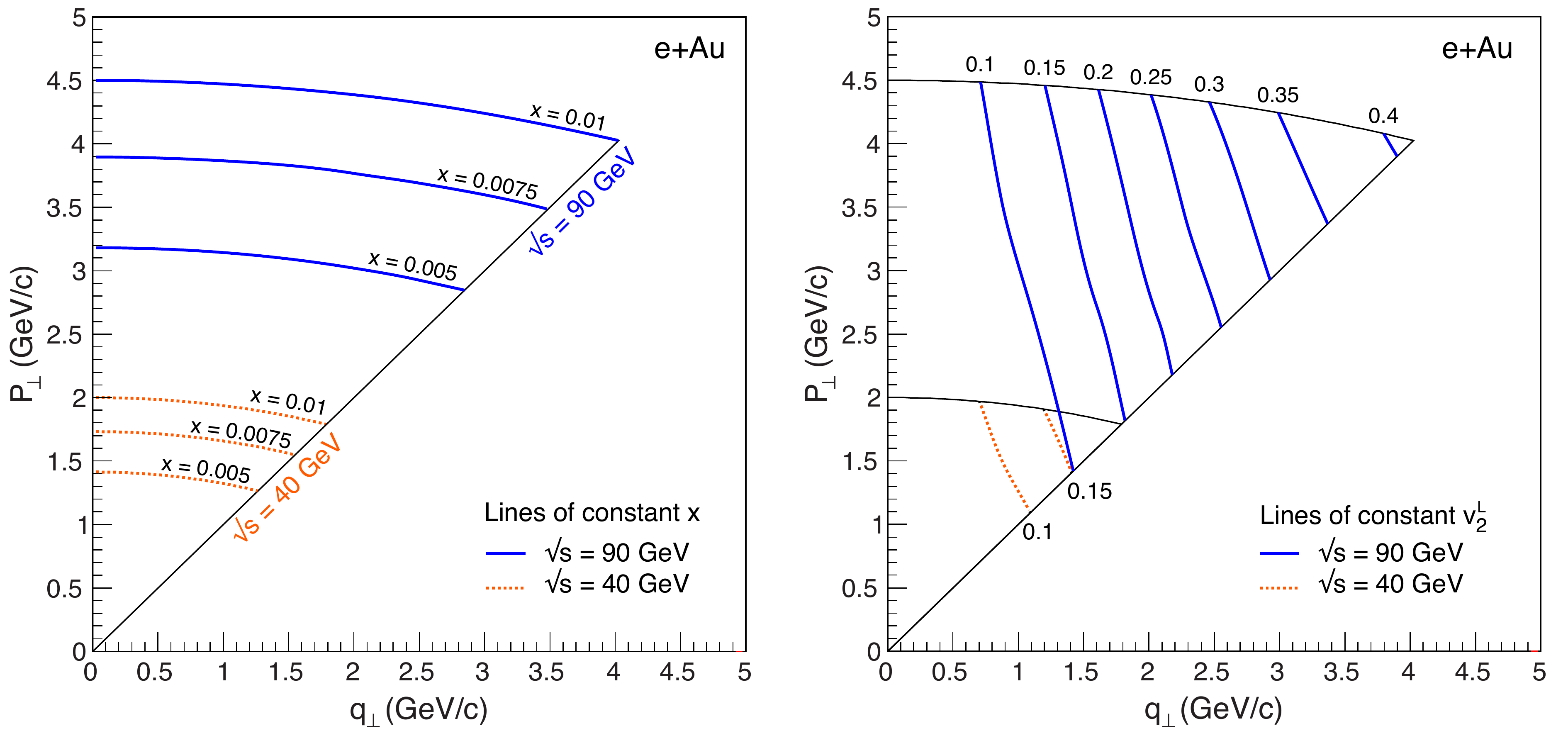}
	\caption{Kinematic range in $q_\perp$ vs.~$P_\perp$ in the
          correlation limit, ${q_\perp} < {P_\perp}$, for two EIC
          energies, $\sqrt{s}$ = 40 and 90 GeV. On the left lines of
          constant $x$ for the respective energies are depicted while
          on the right we show lines of constant azimuthal anisotropy
          for longitudinally polarized virtual photons.}
	\label{fig:KinePieCombo}
\end{figure} 

MCDijet generates a correlated pair of partons per event. It provides
as output the 4-momenta of the two partons, the $z$ value, as well as
general event characteristics such as $W$, $Q^2$, and $x$. Unless
mentioned otherwise we restricted the generation of events to $4 < Q^2
< 90$~GeV$^2$, $x, x_\text{Bj} < 0.01$, $q_\perp > 1$~GeV and $P_\perp >
1.5 q_\perp$. For the ion beam we use Au (A=197).

Figure \ref{fig:KinePieCombo} illustrates the kinematic range in
$q_\perp$ versus $P_\perp$ on the parton level in the relevant region
${q_\perp} < {P_\perp}$, for two EIC energies, $\sqrt{s}$=40 and 90
GeV. In the left plot, we show lines of constant $x$ for both
energies, and on the right, we depict lines of constant azimuthal
anisotropy for longitudinally polarized virtual photons ($v_2^L$). It
becomes immediately clear that substantial anisotropies, $v_2^L \geq
0.15$, can only be observed at the higher energy. Even more important,
from an experimental point of view is the magnitude of the average
transverse momentum $P_\perp$. Jet reconstruction requires
sufficiently large jet energies to be viable. The lower the jet
energy, the more particles in the jet cone fall below the typical
particle tracking thresholds ($p_T\sim 250$ MeV/$c$ in our case),
making jet reconstruction de facto impossible.  For our studies, we
therefore used the highest energy currently discussed for $e$+Au
collisions at an EIC, $\sqrt{s}$ = 90~GeV.

The partons from MCDijet are subsequently passed to parton shower
algorithms from the PYTHIA 8.2 event generator for jet generation. We
assume the dipartons to be $u\bar{u}$ pairs. For jet finding we use
the kt-algorithm from the FastJet package with a cone radius of $R=1$.
In DIS events, jet finding is typically conducted in the Breit frame
(see Sec.~\ref{sec:appB}) which is often seen as a natural choice to
study the final state of a hard scattering.  The Lorentz frame used in
MCDijet is similar to the Breit frame in that the virtual photon and
the proton have zero transverse momenta but distinguishes itself from
the Breit frame by the incoming hadron (Au) beam having the same
energy as in the laboratory frame. Jet finding studies in both frames
showed no significant differences between the two. We therefore used
the ``analysis'' frame described in Sec.~\ref{sec:mcdijet} for all our
studies.

\begin{figure}[htb]
	\centering \includegraphics[width=0.85\linewidth]{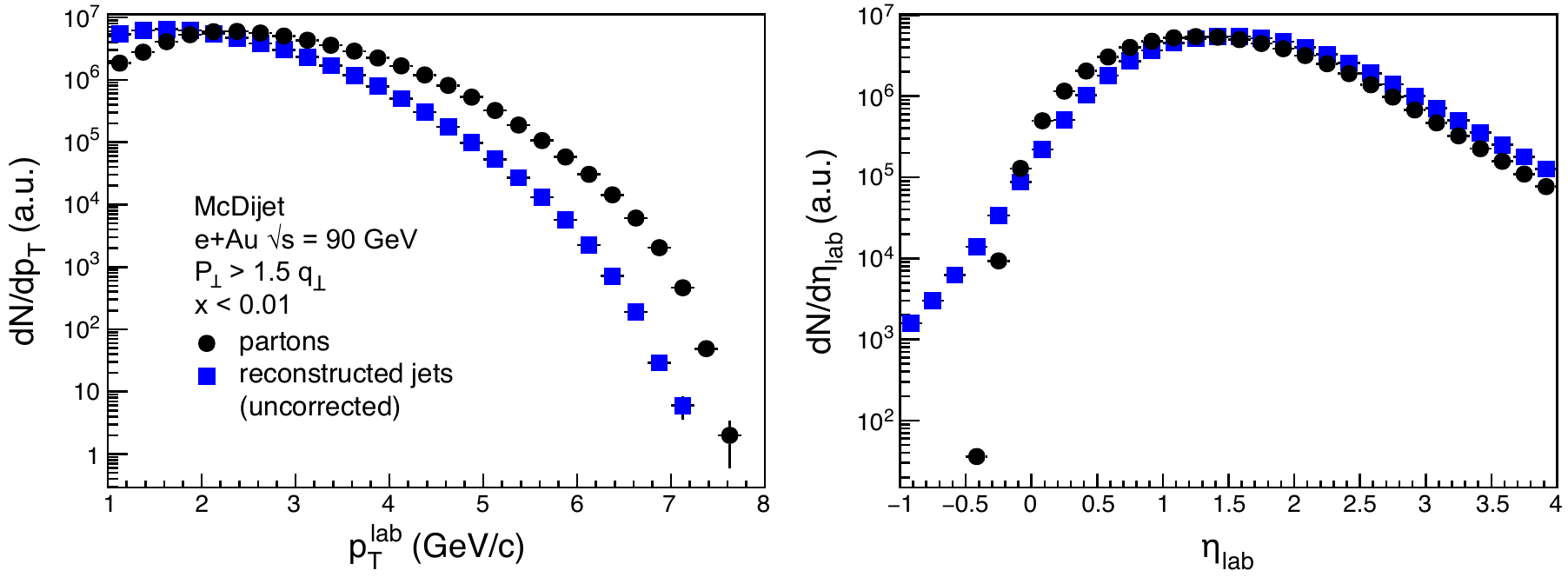}
	\caption{$p_T$ and $\eta$ distributions of partons (filled
          circles) and reconstructed jets (filled squares) in the lab
          frame. The jet spectra are uncorrected.}
	\label{fig:ptetalab}
\end{figure} 
Fig.~\ref{fig:ptetalab} shows the $p_T$ and $\eta$ distributions of
partons (solid circles) from MCDijet and the corresponding reconstructed
jets (solid squares) in the laboratory frame. The uncorrected jet
spectra show the expected shift in $p_T$ due to the loss of particles
below the chosen tracking threshold of 250~MeV/$c$. The pseudorapidity
of the generated partons is well maintained by the jets with a typical
r.m.s.\ of 0.4 units over the whole range. This is caused by unavoidable
imperfections in the jet reconstruction. The smearing becomes more
visible at $\eta < -0.5$ due to the steepness of the spectra.

\begin{figure}[htb]
	\centering \includegraphics[width=0.85\linewidth]{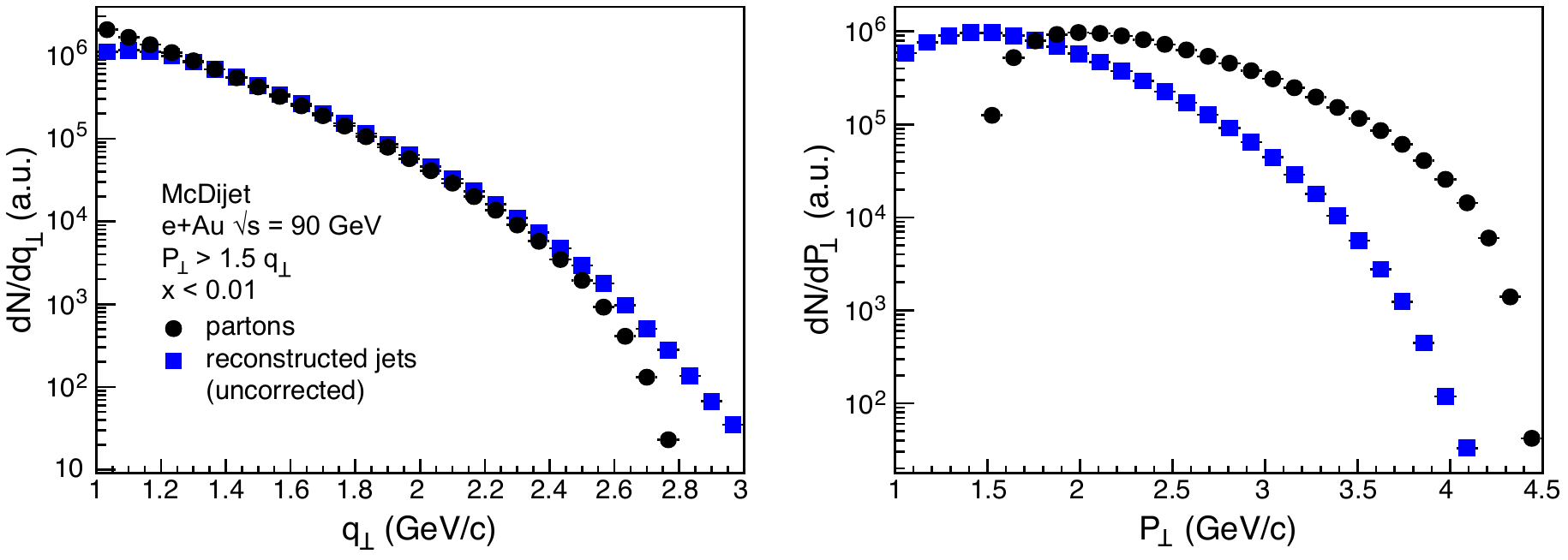}
	\caption{Comparison of $q_\perp$ and $P_\perp$ distribution
          for partons (solid circles) and jets (solid squares).}
	\label{fig:qtPtcm}
\end{figure}
Fig.~\ref{fig:qtPtcm} shows the distribution of events over $q_\perp$
and $P_\perp$. One observes that at the level of reconstructed jets
the distribution over $P_\perp$ is shifted by about $-0.5$~GeV, and
slightly distorted. On the other hand, the distribution in $q_\perp$
of jets reproduces that of the underlying quarks rather accurately,
except for the lowest ($q_\perp \sim 1$~GeV) and highest ($q_\perp
\gsim 2.5$~GeV) transverse momentum imbalances.  In a more in-depth
analysis, which goes beyond the scope of this paper, the jet spectra
would be corrected with sophisticated unfolding procedures (see for
example Refs.~\cite{Schmitt:2016orm, Biondi:2017pzs}). Here, we simply
correct the jet $P_\perp$ spectra by shifting it up so that $\langle
P_\perp \rangle_\mathrm{jet} = \langle P_\perp
\rangle_\mathrm{parton}$ for $P_\perp > 1.5$ GeV/$c$.  No corrections
on $q_\perp$ were applied.

\begin{figure}[htb]
\centering \includegraphics[width=0.85\linewidth]{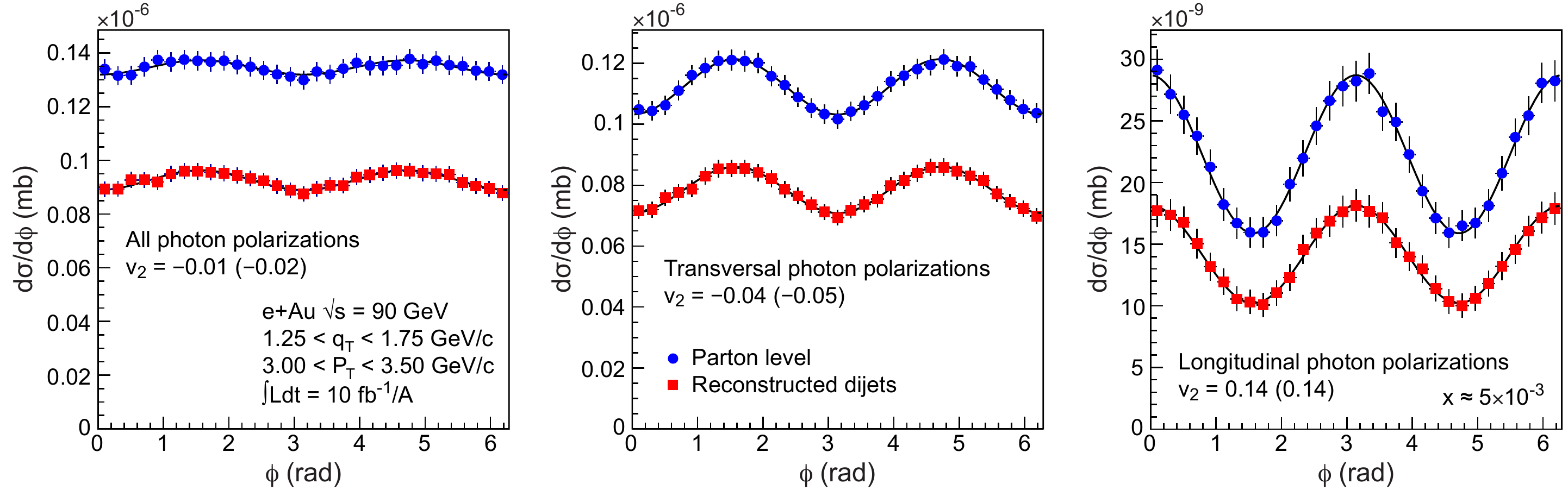}
\caption{$\protect\mathrm{d}\sigma/\protect\mathrm{d}\phi$
  distributions for parton pairs (blue points) generated with the
  MCDijet generator and corresponding reconstructed dijets (red
  points) in $\sqrt{s}$=90 GeV $e$+A collisions for $1.25 < q_\perp <
  1.75$ GeV/$c$ and $3.00 < P_\perp < 3.50$ GeV/$c$. The error bars
  reflect an integrated luminosity of 10~fb$^{-1}$/A. The left plot
  shows the azimuthal anisotropy for all virtual photon polarizations
  while the middle and right plots correspond to transverse and
  longitudinal polarized photons, respectively. For details, see
  text.}
\label{fig:combo}
\end{figure} 
Figure \ref{fig:combo} shows the resulting
$\mathrm{d}\sigma/\mathrm{d}\phi$ distributions for the original
parton pairs (blue solid points) and the reconstructed dijets (red
solid squares) in $\sqrt{s}$=90 GeV $e$+Au collisions for $1.25 <
q_\perp < 1.75$ GeV/$c$ and $3.00 < P_\perp < 3.50$ GeV/$c$. The
results are based on 10M generated events but the error bars were
scaled to reflect an integrated luminosity of 10~fb$^{-1}$/A. The left
plot shows the azimuthal anisotropy for all virtual photon
polarizations, and the middle and right plot for transversal and
longitudinal polarized photons, respectively. The quantitative measure
of the anisotropy, $v_2$, is listed in the figures. The values shown
are those for parton pairs; the accompanying numbers in parenthesis
denote the values derived from the reconstructed dijets. Note the
characteristic phase shift of $\pi/2$ between the anisotropy for
longitudinal versus transversally polarized photons. Despite this
shift, the sum of both polarizations still adds up to nonzero net
$v_2$ due to the dominance of transversely polarized photons, as
depicted in the leftmost plot in Fig.~\ref{fig:combo}.

The reconstructed dijets reflect the original anisotropy at the parton
level remarkably well despite the dijet spectra not being fully
corrected. The loss in dijet yield, mostly due to loss of low-$p_T$
particles, is on the order of $\sim25$\%.  Since the key observable is
the measured anisotropy, the loss in yield is of little
relevance. However, when real data becomes available a careful study
for possible biases will need to be carried out.

\begin{figure}[htb]
	\centering \includegraphics[width=0.85\linewidth]{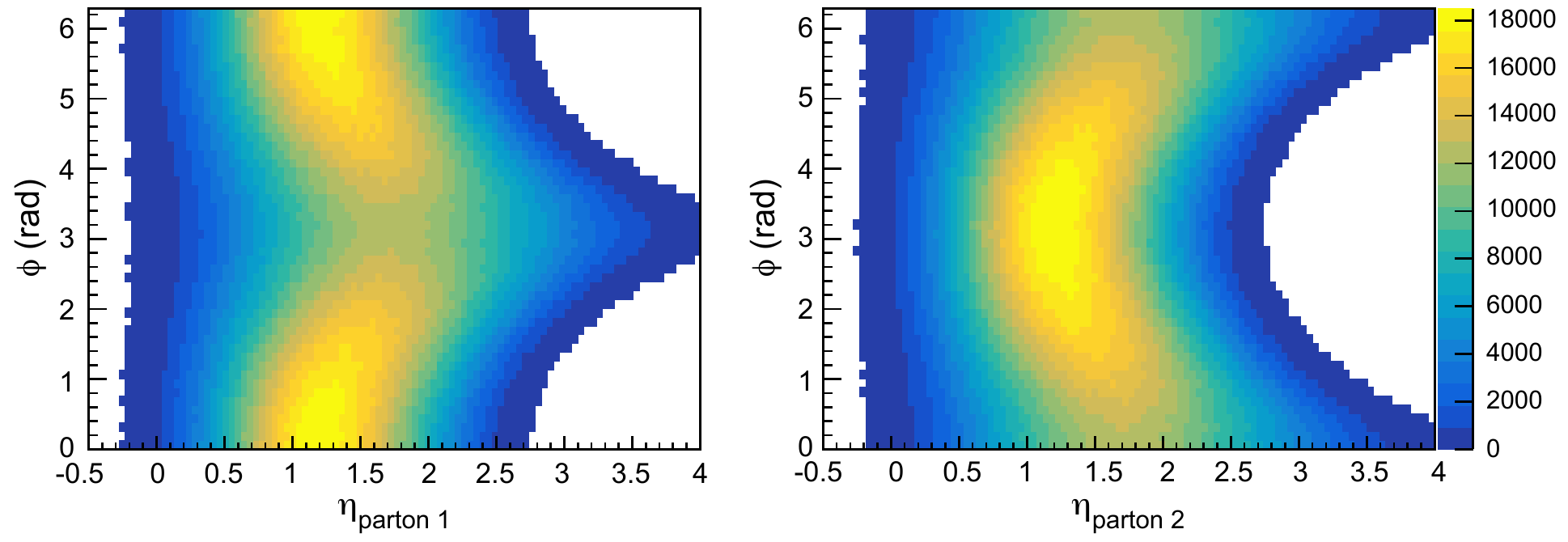}
	\caption{$\phi$, the angle between $\vec
          P_\perp$ and $\vec q_\perp$,  as a function of the
          pseudorapidity, $\eta$, of each of the partons. The strong
          correlation indicates the sensitivity of the observed
          anisotropy in $\phi$ on the $\eta$ acceptance of potential
          experimental measurements.}
	\label{fig:butterfly}
\end{figure} 
In our studies we noted the momentous correlation of the angle $\phi$
with the pseudorapidity, $\eta$, of the partons/jets illustrated in
Fig.~\ref{fig:butterfly}. This behavior is introduced through the
$\eta$ dependence of $z$ and can be illustrated by expressing $z$
through the kinematics of the two partons as:
\begin{equation}
	z = \frac{E_1 + k_{1z}} {(E_1 + k_{1z})+(E_2 + k_{2z} )}\,,
\end{equation}
where $k_{i z } = E_i \tanh \eta_i = k_{i \perp} \sinh \eta_i$.
Recall that $z$ is the momentum fraction of the first and $1-z$ that
of the second parton/jet.  Rewriting $P_\perp$ (see
Eq.~\ref{eq:Ptqtdef}) as $\vec{P}_{\perp } = \vec{k}_{1\perp} - z \vec
q_\perp$ shows that for $z \rightarrow 1$ large $P_\perp$ are
biased towards $\vec P_\perp \uparrow\downarrow {\vec q_\perp}$ thus
favoring $\phi \approx \pi$.  On the other hand, writing
$\vec{P}_{\perp } = -\vec{k}_{2\perp} + (1-z) \vec q_\perp$
we see that for $z \rightarrow 0$ large
$|\vec{P}|$ prefers $\vec{P}_{\perp } \uparrow\uparrow
\vec q_\perp$, i.e.\ $\phi \approx 0~\text{mod}~2\pi$. This has
substantial impact on the experimental measurement since even in the
absence of any anisotropy the finite $\eta$ acceptance of tracking
detectors will generate a finite and positive $v_2$. On the other
hand, a tight rapidity range also alters the actual anisotropy. For
example the generated $v_2^L$ anisotropy in the right plot of
Fig.~\ref{fig:combo} of 14\% requires at a minimum a range of $0 <
\eta < 3$; for $0 < \eta < 1.5$ the observed $v_2^L$ shrinks to $\sim
0.05$.  This effect was verified with PYTHIA simulations where a
limited acceptance showed a considerable effect despite PYTHIA having
no mechanism to generate any intrinsic anisotropy. Only for wide
acceptances with $\Delta\eta \geq 3$ does the $\phi$ distribution
become flat. Measurements at an EIC will need to be corrected for
these massive finite acceptance effects.


\subsection{Background studies}
\label{subsec:SB}

While MCDijet allows the study of the signal anisotropy in great
detail it does neither generate complete events, nor does it allow us
to derive the level of false identification of dijets in events
unrelated to dijet production. The purity of the extracted signal
sample ultimately determines if these measurements can be conducted.
For studies of this kind we have to turn to PYTHIA6, an event
generator that includes a relatively complete set of DIS processes.

\begin{figure}[htb]
\centering \includegraphics[width=0.25\linewidth]{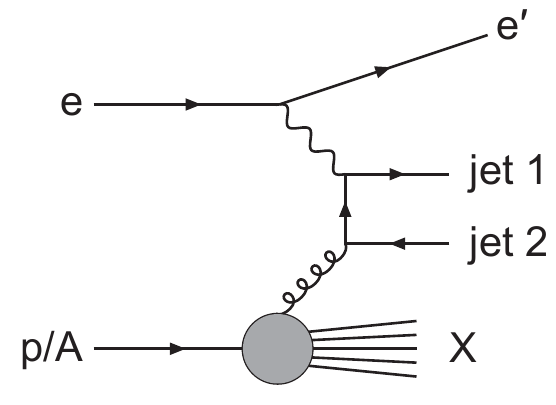}
\caption{Photon-gluon fusion processes that contributes to the 2+1 jet signal cross-section.}
\label{fig:diagram}
\end{figure}
The presence of underlying event activity is key to answering the
question if one can achieve a clear separation between the products of
the hard partonic interaction and the beam remnants. For that reason,
one usually labels an event as ``2+1 jets'' if it has 2 jets coming
from the hard partonic interaction, with the ``+1'' indicating the
beam remnants.  The diagram in Fig.~\ref{fig:diagram} thus depicts a
2+1-jet event.

While dijet studies have been successfully conducted in $e+p$
collisions at HERA (see for example \cite{Aktas:2007bv,
  Gouzevitch:2008zza}) most such measurement have been carried out at
high $Q^2$ and high jet energies ($E_\mathrm{jet} > 10$~GeV). In our
studies, however, we focus on moderately low virtualities and
relatively small jet transverse momenta $P_\perp$ (see
Fig. \ref{fig:ptetalab}). Consequently, the dijet signal is easily
contaminated by beam remnants.  To minimize this background source we
limit jet reconstruction to $1 < \eta < 2.5$, sufficiently far away from
the beam fragmentation region.

\begin{figure}[htb]
\centering \includegraphics[width=0.5\linewidth]{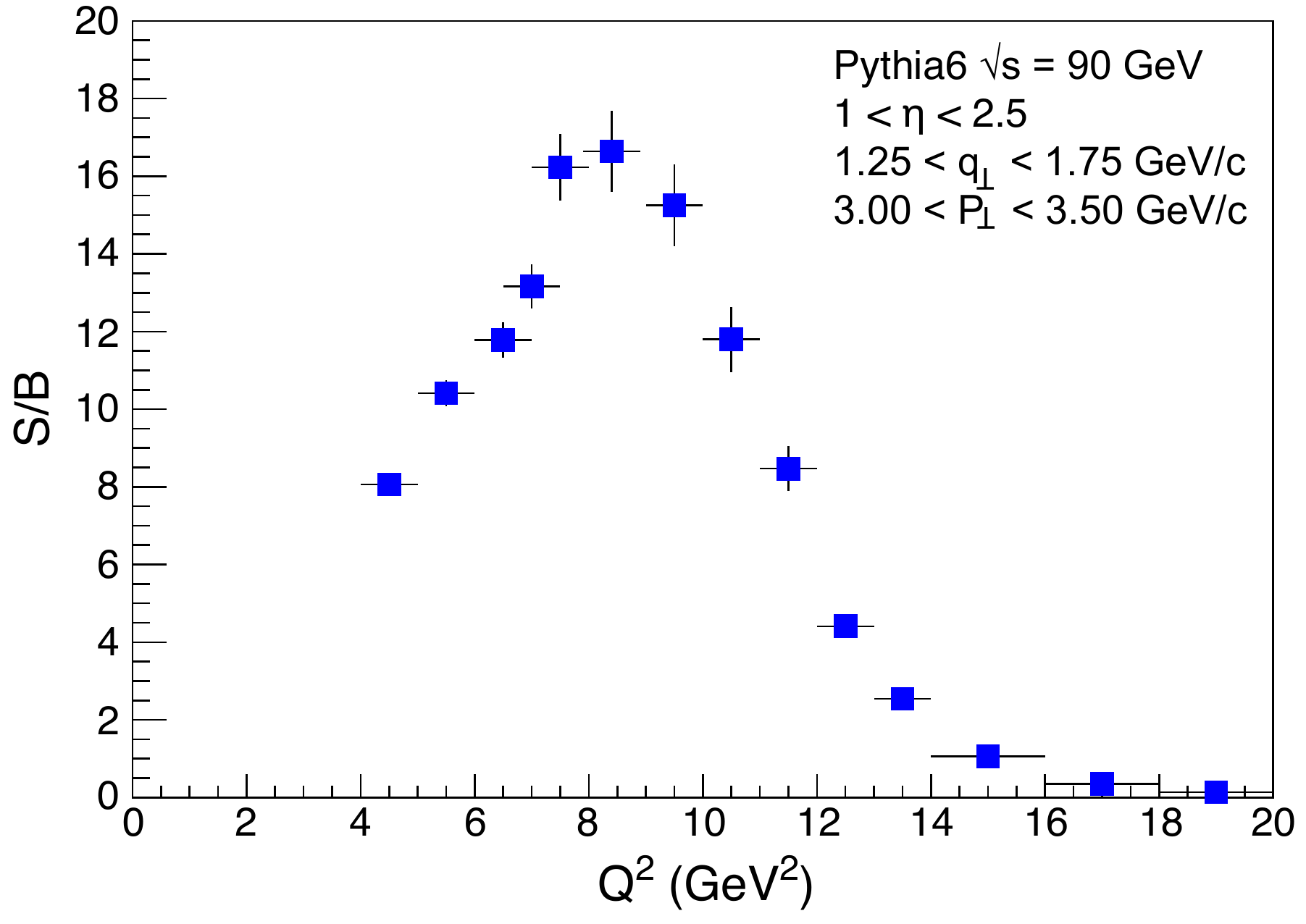}
\caption{$Q^2$ dependence of the signal-to-background ratio derived from PYTHIA6.}
\label{fig:Q2Dep}
\end{figure} 
In our PYTHIA6 study we count $f_i + \gamma^*_\mathrm{T,L} \rightarrow
f_i + g$ and $g + \gamma^*_\mathrm{T,L} \rightarrow f_i + \bar{f_i}$
(see Fig.~\ref{fig:diagram}) as signal and all other as
background processes. By far the dominant background source is the
standard LO DIS process $ \gamma^* + q \rightarrow q$. Figure
\ref{fig:Q2Dep} illustrates the $Q^2$ dependence of the
signal-to-background (S/B) ratio, {\it i.e.}, the number of correctly
reconstructed signal events over the number of events that were
incorrectly flagged as containing a signal dijet process. The S/B
ratio rises initially due to the improved dijet reconstruction
efficiency towards larger $Q^2$ (or $P_\perp$) but then drops
dramatically as particles from the beam remnant increasingly affect
the jet finding. In what follows we therefore limit our study to $4
\leq Q^2 \leq 12$ GeV$^2$.

\begin{figure}[htb]
	\centering \includegraphics[width=0.5\linewidth]{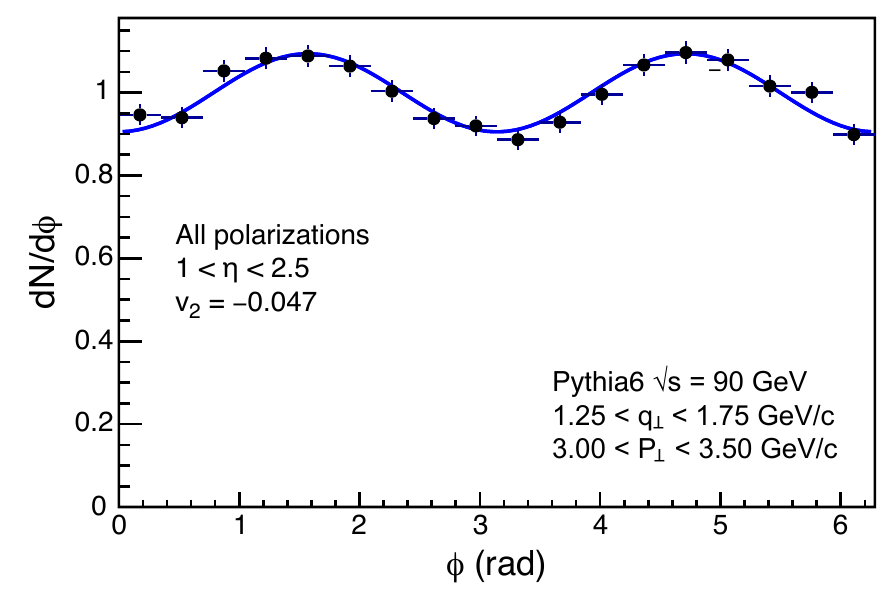}
	\caption{Azimuthal asymmetry in reconstructed dijet events from PYTHIA caused by the  limited $\eta$  acceptance.}
	\label{fig:corr}
\end{figure} 
As discussed in Sec.~\ref{sec:mcdijetStudy}, the necessity to limit
dijet reconstruction to $\eta < 2.5$ creates a substantial anisotropy
illustrated in Fig.~\ref{fig:corr}. The corresponding $v_2$ is always
negative regardless of the polarization of the virtual photon and
different from the true signal where $v_2^L$ and $v_2^T$ have
opposite signs. This is a plain artifact of the limited
pseudorapidity range. For a wider $\eta$ range the modulation vanishes
but the S/B drops substantially since beam fragmentation remnants
start to leak in. Since the anisotropy is of plain kinematic origin it
can be easily derived from Monte-Carlo and corrected for. In the
following we subtracted this $\eta$-range effect from our data sample.

\begin{figure}[htb]
	\centering \includegraphics[width=0.5\linewidth]{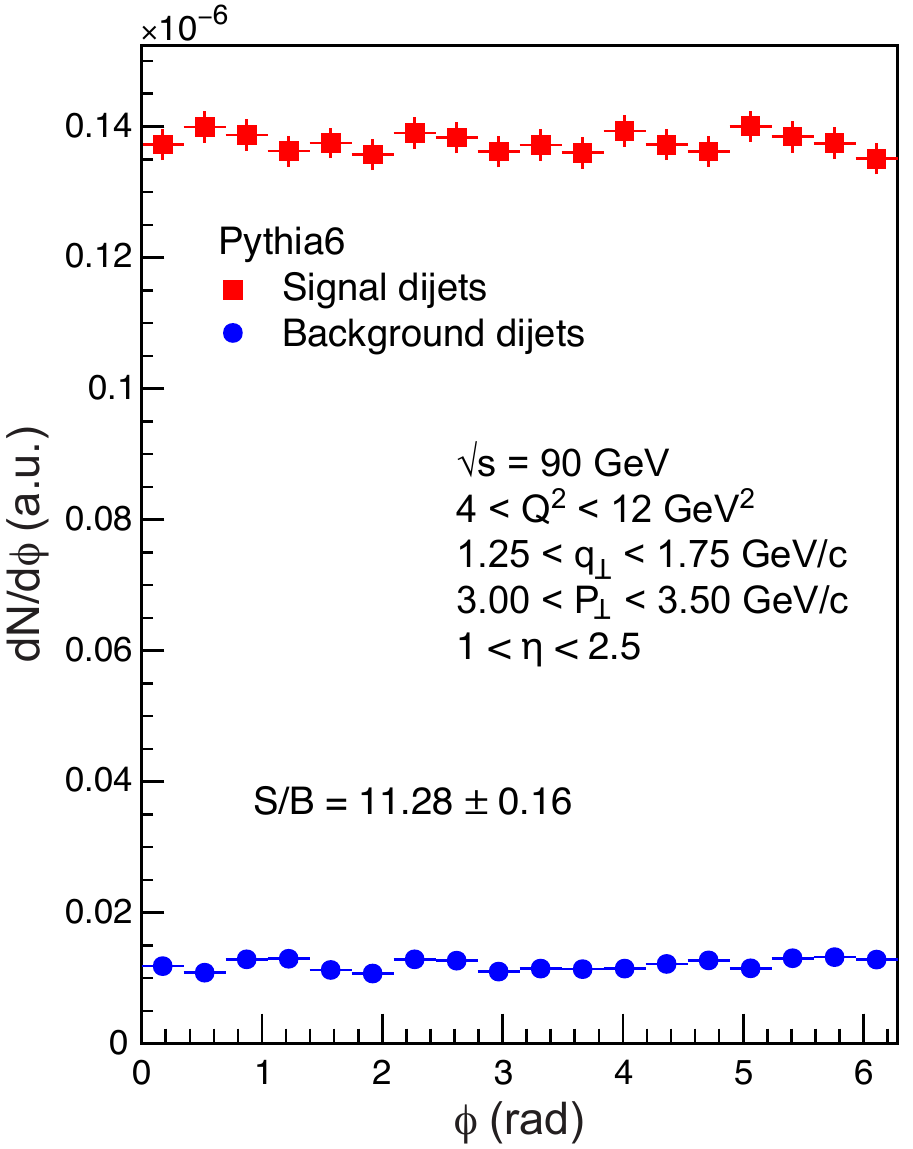}
	\caption{$dN/d\phi$ distribution of signal and background jets
          after corrections.}
	\label{fig:SB}
\end{figure} 
Figure \ref{fig:SB} shows the resulting $dN/d\phi$ distributions for
signal jets (solid squares) and background jets (solid circles). The
signal-to-background ratio for the indicated cuts is S/B $\approx$ 11.
After the finite $\eta$-range correction both, signal and background
pairs show no modulation, as expected.

\subsection{Extracting $v_2^L$ and $v_2^T$}

In order to derive the distribution of linearly polarized gluons via
Eqs.~(\ref{eq:v2_L_T}), the contributions from transverse ($v_2^T$)
and longitudinally polarized photons ($v_2^L$) need to be
disentangled. With the exception of diffractive $J/\psi$ production,
no processes in DIS exist where the polarization of the virtual photon
can be measured directly. In our case there are 3 features that do
make the separation possible: (i) $v_2^L$ and $v_2^T$ have opposite
signs (see Fig.~\ref{fig:combo}), (ii) the background contribution
shows no anisotropy (see Fig.~\ref{fig:SB}), and (iii) the relation
\begin{equation}
	v_2^\mathrm{unpol} = \frac{R v_2^L + v_2^T}{1+R}
	\label{Eq:v2sum}
\end{equation}
ties together the
unpolarized, {\it i.e.} measured, $v_2$ with the transverse and
longitudinal components.  $R$ is a kinematic factor depending entirely
on known and measured quantities~\footnote{The expression for $R$ is derived
  from the leading order cross-sections~(\ref{eq:dijet_T},
  \ref{eq:dijet_L}).}:
\begin{equation}
R =
\frac{8 \epsilon_f^2 P_\perp^2\, z (1-z) }{(z^2+(1-z)^2) \, (\epsilon_f^4 +
  P_\perp^4)}\,.
\label{eq:Rformula}
\end{equation}

\begin{figure}[htb]
	\centering \includegraphics[width=0.65\linewidth]{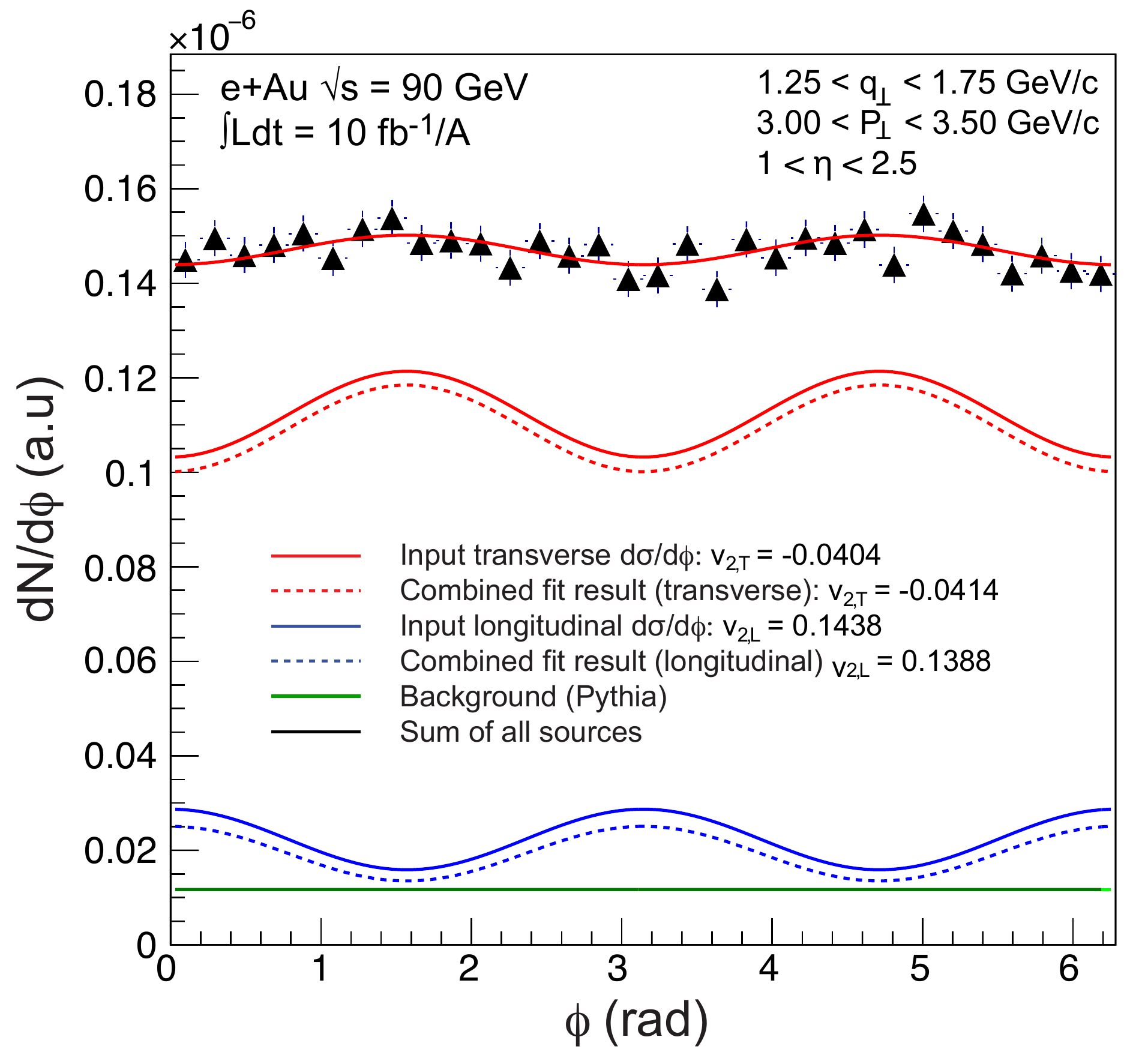}
	\caption{Result of a fit of combined signal and background to a data sample obtained
		in $\sqrt{s}=90$ GeV $e$+A collisions with an integrated luminosity of 10 fb$^{-1}$/A. For details see text.}
	\label{fig:extract}
\end{figure} 
Our strategy is to perform a combined 5-parameter fit of all 3
components to the full data sample: The signal for longitudinal polarization
($\sigma_L, v_2^L$), that for transverse polarization ($\sigma_T,
v_2^T$), and the flat background ($\sigma_b$). The fit uses
the constraint provided by Eq.~\eqref{eq:Rformula}. We generated the
data sample in a separate Monte-Carlo combining the signal from
MCDijet with the background contribution from PYTHIA6 while smearing
each data point randomly according to the statistics available at a
given integrated luminosity. The fit provides the desired $v_2^L$ and
$v_2^T$. In order to determine the corresponding errors we repeat the fit
10,000 times and derive the standard deviation from the obtained
$dN/dv_2^{L,T}$ distributions.  With reasonable accuracy the errors
are distributed symmetrically about the true value.

Figure \ref{fig:extract} shows the result of one typical fit on data
generated for a integrated luminosity of 10 fb${^1}$/A. The scatter
and errors on the data points reflect the size of the potential data
sample, the red and the blue curves illustrate the input (solid curve)
and the fit result (dashed curve) for $v_2^L$ and $v_2^T$. The dashed
curves were offset for better visibility.

\renewcommand{\arraystretch}{0.5}
\begin{table}
	\begin{tabular}{|c | c | c |} \hline
		Integrated Luminosity (fb$^{-1}$/A) & $\delta v_2^L/v_2^L$ (\%) &  $\delta v_2^T/ v_2^T (\%)$ \\ \hline \hline
		1 & 23.7 & 16.7 \\  
		10 & 7.5 & 5.3 \\  
		20 & 5.5 & 3.9 \\ 
		50 & 3.4 & 2.4 \\
		100 & 2.4 & 1.6 \\ \hline
	\end{tabular}
\caption{Relative error on the extracted $v_2^L$ and $v_2^T$ for various integrated luminosities.}
\label{tab:errors}
\end{table}
Table \ref{tab:errors} shows the derived relative errors on $v_2^L$
and $v_2^T$ for various integrated luminosities. These listed
uncertainties refer only to the selected cuts of $1.25 < q_\perp <
1.75$ GeV/$c$, $3 < P_\perp < 3.5$ GeV/$c$, $4 < Q^2 < 12$ GeV$^2$,
and $1 < \eta < 2.5$. The errors show the expected $({\int
  {\cal{L}} \mathrm{d}t})^{-1/2}$ scaling. Systematic studies showed that the
relative errors improve with increasing $P_\perp$, {\it i.e.},
increasing $v_2$.  Our results indicate that a proper measurement of
the linearly polarized gluon distribution will require integrated
luminosities of at least 20 fb$^{-1}$/A or more. Hence, this
measurement would be a multi-year program assuming that an EIC
initially starts off with luminosities around $10^{33}$ cm$^{-2}$
s$^{-1}$. The errors were derived assuming cross-section generated by 
MCDijet that are, as described earlier, somewhat lower than the ones
from PYTHIA6. 

\FloatBarrier

\section{Outlook}
\label{sec:outlook}

Our current proof of principle analysis relied on a variety of
simplifications and approximations as our main focus was on the
reconstruction of relatively low $p_T$ jets and their angular
distribution. In this section we address some improvements that 
would improve the accuracy of the model and of the analysis.

First off, a more realistic modelling of the impact parameter
dependence of the thickness of the target nucleus would be
useful. This is due to the fact that cuts on the minimal $P_\perp$
introduce a bias towards more central impact parameters as the dijet
cross-section decreases with $P_\perp$ but increases with the
saturation scale $Q_s$. In fact, this bias does also affect the shape
of the small-$x$ gluon distributions as functions of
$q_\perp/Q_s$~\cite{Dumitru:2017ftq,Dumitru:2017cwt,Dumitru:2018iko}. To
account for this effect the event generator would have to employ
individual JIMWLK field configurations rather than the unbiased
average distributions.

Another improvement is to include running coupling corrections to the
dijet cross-section and to small-$x$ JIMWLK
evolution~\cite{Lappi:2012vw}. These would be important, in
particular, if the analysis is performed over a broad range of
transverse momenta.

The measurement of the distribution of linearly polarized gluons via
the $\cos 2\phi$ azimuthal dependence requires significant jet
momentum imbalance $q_\perp$ not much less than the saturation scale
$Q_s$. On the other hand, the cross-section decreases steeply with
$P_\perp$ and so, in practice, the ratio $q_\perp/P_\perp$ cannot be
very small. Hence, power corrections to Eqs.~(\ref{eq:dijet_T},
\ref{eq:dijet_L}) may be significant and should be
implemented. (Expressions for the leading power corrections in the
large-$N_c$ limit can be found in Ref.~\cite{Dumitru:2016jku}).

One should also account for the Sudakov suppression which arises due
to the presence of the two scales $q_\perp$ and
$P_\perp$~\cite{Mueller:2013wwa, Zheng:2014vka, Boer:2017xpy}. In view
of the relatively large ratio of $q_\perp/P_\perp$ employed in our
analysis we do not expect a very large suppression of the amplitude of
the $\cos 2\phi$ azimuthal dependence.

Given that the light-cone momentum fraction of the target partons is
not very small even at the highest energies envisaged for an EIC it
would be important to also account for the $\gamma^* + q \to g+q$
process, unless one attempts to identify events producing a gluon jet.

As the Electron-Ion Collider projects progresses detector concepts will become  more refined. Once the design of the envisioned multi-purpose detector(s) are finalized the feasibility study  discussed in this paper should
be repeated using detailed detector effects (acceptance, resolution) 
and include full unfolding procedures that would improve over the
simple corrections used in this work. There is an increasing interest in 
jet studies at an EIC that could potentially lead to improved jet finding
procedures tailored to the specific kinematics and energies relevant for this study.

\section{Summary and conclusions}
\label{sec:concl}

This paper presents a study of the feasibility of measuring the
conventional and linearly polarized Weizs\"acker-Williams (WW) transverse
momentum dependent (TMD) gluon distributions at a future high-energy
electron-ion collider via dijet production in Deeply Inelastic
Scattering on protons and nuclei at small $x$. In particular, we have found that
suitable cuts in rapidity allow for a reliable separation of the dijet
produced in the hard process from beam jet remnants. A cut on the
photon virtuality, $Q^2 \lsim 12$~GeV$^2$ suppresses the LO
$\gamma^* q \to q+X$ process and leads to a signal to background ratio
of order 10.

The jet transverse momentum $\vec P_\perp$ as well as the momentum
imbalance $\vec q_\perp$, and the azimuthal angle $\phi$ between these
vectors can all be reconstructed with reasonable accuracy even when
$P_\perp$ is on the order of a few GeV. The $\phi$-averaged dijet
cross-section determines the conventional WW TMD
$xG^{(1)}(x,q^2_\perp)$ while $v_2=\langle\cos 2\phi\rangle \sim
xh^{(1)}_\perp(x,q^2_\perp) \, /\, xG^{(1)}(x,q^2_\perp)$ is
proportional to the ratio of the linearly polarized to conventional WW
TMDs. Furthermore, with known $P_\perp$, $Q^2$ and jet light cone
momentum fraction $z$ it is possible to separate $v_2$ into the
contributions from longitudinally or transversely polarized photons,
respectively, to test the predicted sign flip, $v_2^L\cdot v_2^T <
0$. We estimate that with an integrated luminosity of $20$~fb$^{-1}/A$
one can determine $v_2^L$ and $v_2^T$ with a statistical error of
approximately 5\%.

\appendix 

\section{Breit frame}
\label{sec:appB}

In any frame the ratio of plus momenta of quark and
virtual photon  is given by
\begin{equation}
	z = \frac{k_1^+}{q^+}  = \frac{|\vec{k}_1| + k_{1z}}{q_0 + q_z}~,
\end{equation}
and therefore  in any frame
\begin{equation}
	k_{1z} = \frac{ [ z (E_0 + q_z) ]^2 - k_{1\perp}^2} { 2 z (q_0
	+ q_z)} ~.
\end{equation}
Similarly, for the antiquark
\begin{equation}
	k_{2z} = \frac{ [ \bar{z} (E_0 + q_z) ]^2 - k_{2\perp}^2} { 2 \bar{z} (q_0 + q_z)}
\end{equation}

In particular, in the Breit frame ($q_0=0$ and $|q_z|=Q$) we get
\begin{equation}
	k_{1z} = \frac{ ( z  Q )^2 - k_{1\perp}^2} { 2 z  q_z}\, . 
\end{equation}
Taking the longitudinal momentum of the photon to be positive
(following the convention in the MCDijet code),
\begin{equation}
	k_{1z} = \frac{ ( z  Q )^2 - k_{1\perp}^2} { 2 z  Q}\,.
\end{equation}
Recalling that $\bar{z} = 1-z$ we can finally write the longitudinal
momenta of the quark and anti-quark in the Breit frame in the form
\begin{eqnarray}
	k_{1z} &=& \frac{ ( z  Q )^2 - k_{1\perp}^2} { 2 z  Q}  \\
	k_{2z} &=& \frac{ [ (1-z)  Q ]^2 - k_{2\perp}^2} { 2 (1-z)  Q}~.
\end{eqnarray}
The longitudinal boost leading from
Eqs.~(\ref{eq:k1z_analysis-frame},\ref{eq:k2z_analysis-frame}) to
these expressions defines the transformation from our ``analysis
frame'' to the Breit frame.

\begin{acknowledgments}
We thank
Elke-Caroline Aschenauer,
Jin Huang,
Larry McLerran,
Tuomas Lappi,
Elena Petreska,
Andrey Tarasov,
Prithwish Tribedy,
Pia Zurita
for useful discussions.

A.D. gratefully acknowledges support by the DOE Office of Nuclear Physics through Grant No. DE-FG02- 09ER41620; and from The City University of New York through the PSC-CUNY Research grant 60262-0048.

V.S. thanks the ExtreMe Matter Institute EMMI (GSI Helmholtzzentrum f\"ur Schwerionenforschung, Darmstadt, Germany)  for partial support and their hospitality.

T.U.'s work was supported by the Office of Nuclear Physics within the U.S. DOE Office of Science.
\end{acknowledgments}

\bibliography{bibliography}

\end{document}